\documentclass[twoside, a4paper, notoc, 10pt]{JHEP3}
\usepackage{amsfonts}
\usepackage{amssymb}
\usepackage{comment}
\usepackage{epsfig}
\usepackage{graphicx}
\usepackage{amsmath}
\usepackage{amsxtra}

\newcommand{\bea}{\begin{eqnarray}}
\newcommand{\eea}{\end{eqnarray}}
\newcommand{\vo}{{\cal V}}
\newcommand{\be}{\begin{equation}}
\newcommand{\ee}{\end{equation}}
\newcommand{\mc}{\mathcal}
\newcommand{\mbb}{\mathbb}

\def\ba{\begin{eqnarray}}
\def\ea{\end{eqnarray}}
\def\nn{\nonumber}

\makeatletter
\def\x@arrow{\DOTSB\Relbar}
\def\xlongequalsignfill@{\arrowfill@\x@arrow\Relbar\x@arrow}
\newcommand{\xlongequal}[2]{%
    \ext@arrow 0099\xlongequalsignfill@{#1}{#2}}
\makeatother

\newcommand{\roughly}[1]{\mathrel{\raise.3ex\hbox{$#1$\kern-0.85em
\lower1ex\hbox{$\sim$}}}}

\def\endignore{}
\def\ignore #1\endignore{}
\def\nn{\nonumber}

\def\beq{\begin{equation}}
\def\eeq{\end{equation}}
\def\beqa{\begin{eqnarray}}
\def\eeqa{\end{eqnarray}}

\newcommand{\bmat}{\left(\begin{array}}
\newcommand{\emat}{\end{array}\right)}

\def\endignore{}
\def\ignore #1\endignore{}

\def\-{\hphantom{-}}

\def\s2{\frac{1}{2}}

\def\IF{\relax{\rm I\kern-.18em F}}
\def\II{\relax{\rm I\kern-.18em I}}
\def\IP{\relax{\rm I\kern-.18em P}}
\def\IC{\relax{\rm I\kern-.48em C}}
\def\IR{\relax{\rm I\kern-.18em R}}
\def\IK{\relax{\rm I\kern-.20em K}}
\def\IM{\relax{\rm I\kern-.25em M}}

\def\Dsl{\,\raise.15ex\hbox{/}\mkern-13.5mu D}

\def \one{\relax{\rm 1\kern-.26em I}}

\def\nn{\nonumber}

\def\({\left(}
\def\){\right)}

\title{De Sitter String Vacua from Dilaton-dependent Non-perturbative Effects}

\author{Michele Cicoli,${}^{1,2}$ Anshuman Maharana,${}^3$ F.~Quevedo${}^{1,3}$ and C.~P.~Burgess${}^{4,5}$ \\

$^1$ Abdus Salam ICTP, Strada Costiera 11, Trieste 34014, Italy

$^2$ INFN, Sezione di Trieste, Italy

$^3$ DAMTP/CMS, University of Cambridge, \\ \qquad Wilberforce Road,
 Cambridge CB3 0WA, UK.

$^4$ Department of Physics \& Astronomy, McMaster University\\ \qquad 1280 Main Street West, Hamilton ON, Canada.

$^5$ Perimeter Institute for Theoretical Physics\\
\qquad 31 Caroline Street North, Waterloo ON, Canada.

}

\abstract{We consider a novel scenario for modulus stabilisation in IIB string compactifications in which the K\"ahler moduli are stabilised by a general set-up with two kinds of non-perturbative effects: $(i)$ standard K\"ahler moduli-dependent non-perturbative effects from gaugino condensation on D7-branes or E3-instantons wrapping four-cycles in the geometric regime; $(ii)$ dilaton-dependent non-perturbative effects from gaugino condensation on space-time filling D3-branes or E(-1)-instantons at singularities. For the LARGE Volume Scenario (LVS), the new dilaton-dependent non-perturbative effects provide a positive definite contribution to the scalar potential that can be arbitrarily tuned from fluxes to give rise to de Sitter vacua. Contrary to anti D3-branes at warped throats, this term arises from a manifestly supersymmetric effective action. In this new scenario the `uplifting' term comes from F-terms of blow-up modes resolving the singularity of the non-perturbative quiver. We discuss phenomenological and cosmological implications of this mechanism. This set-up also allows a realisation of the LVS for manifolds with zero or positive Euler number. \\

{$e$-mail: \email{mcicoli@ictp.it}; \email{A.Maharana@damtp.cam.ac.uk}  \\
\email{F.Quevedo@damtp.cam.ac.uk}; \email{cburgess@perimeterinstitute.ca}}}

\preprint{DAMTP-2012-7}

\begin{document}

\tableofcontents

\bigskip

\section{Introduction}

Non-perturbative effects play a very important r\^ole in the study of low-energy implications of string theory. In the past few years a very good understanding of these effects has been achieved despite the absence of a fully non-perturbative formulation of the theory. Much progress has been made on the origin, structure and computability of these terms in the effective field theory at low energy and weak coupling. D-branes and their Euclidean counterparts generalise the r\^ole of non-perturbative effects (gaugino condensation and instantons) in field theories.

In IIB string compactifications to four dimensions with $\mc{N}=1$ supersymmetry and O3/O7-planes,
there are two known ways to generate non-perturbative effects \cite{NPreview}:
\begin{itemize}
\item by gaugino condensation in the field theory living on D7-branes or Euclidean D3-brane (E3) instantons wrapped on geometric four-cycles;
\item by gaugino condensation on spacetime-filling D3-branes or Euclidean D(-1)-brane (E(-1)) instantons at singularities.
\end{itemize}
The corresponding gauge kinetic function is determined in the first case by the K\"ahler modulus $T$ controlling the size of the wrapped four-cycle whereas in the second case by the dilaton field $S$ and the blow-up mode $Q$ resolving the singularity.

Despite being very much suppressed at weak coupling,
non-perturbative effects are nevertheless relevant as corrections to the superpotential $W$ due to non-renormalisation theorems that keep nominally higher-order contributions from contributing. For instance, symmetries forbid the appearance of the K\"ahler moduli in $W$ at any order in perturbation theory and so their appearance in $W$ is necessarily purely non-perturbative. For the simpler constructions normally considered the non-perturbative superpotential coming from these two types of mechanisms can be written as:
\be \label{vanillaWnp}
W_{\rm np} =\sum_i A_i\, e^{-\,a_i \, \left( T_i+ h_i(\mc{F}) S\right)} + B\, e^{-\, b \, \left(S+ h_j\, Q_j\right)}\,,
\ee
where the prefactors $A_i$ and $B$ are in general functions of the complex structure moduli and the open string moduli (D7-deformation and D3-position moduli). The constants $a_i$ and $b$ are given by $a_i=2\pi/N_i$ and $b=2\pi/N_b$ where, in the case of gaugino condensation, $N_i=\beta_i/3$ and $N_b=\beta_b/3$ with $\beta_i$ and $\beta_b$ the one-loop beta function coefficients of the condensing gauge theories, while, in the case of instantons, $N_i=N_b=1$.

The quantities $h_i(\mc{F})$ and $h_j$ are non-zero in the presence of a non-vanishing magnetic flux $\mc{F}$
on the world-volume of a D7-brane or if a stack of D3-branes is placed at a singularity. In this way chiral matter is generated in the low-energy theory.
More precisely, for flux on a D7-brane the K\"ahler moduli get charged under the anomalous $U(1)$ and the gauge kinetic function $f_i$ gets a flux-dependent shift $h_i(\mc{F})$ that introduces a new dependence on the dilaton, modifying it from $f_i=T_i$ to $f_i=T_i+h(\mc{F}_i) S$. On the other hand, for D3-branes at a singularity the blow-up modes $Q_j$ resolving the singularity get charged under the anomalous $U(1)$ and the gauge kinetic function receives extra contributions of the form $f=S + h_j Q_j$ where $h_j$ are constants proportional to the $U(1)$-charge of the blow-up modes. In both cases the coefficients $A_i$ and $B$ must also depend on charged matter fields in order for $W$ to be gauge invariant.

To date only the $T_i$ dependent part in $W_{\rm np}$ has been used for modulus stabilisation. However, there is no reason to assume $B=0$ and therefore also the $S$-dependent term in $W_{\rm np}$ has to be included unless it can be argued to be subdominant. One of the main points of this article is to explore the implications of these non-perturbative effects for modulus stabilisation.

We here study their inclusion within the context of the LARGE Volume Scenario (LVS) \cite{LVS}, which provides a concrete and controlled realisation of modulus stabilisation in flux compactifications of IIB string theory. The mechanism used in the LVS is very generic \cite{ccq2} and shows that the natural minimum of the leading contributions to the scalar potential corresponds to (non-supersymmetric) anti de Sitter (AdS) 4D spacetime with exponentially large volume. As for other stabilisation scenarios it is necessary to seek uplifting mechanisms that can lift the minimum to flat or de Sitter (dS) vacua, and it has been assumed that this can be realised in a similar manner as in the KKLT scenario \cite{KKLT}, where an anti D3-brane is located at the tip of a highly warped throat in the extra dimensions. The anti-brane provides a positive-definite contribution to the scalar potential and the warping can be adjusted to explicitly tune the minimum
to have a small positive vacuum energy.

The use of anti-branes in this way raises standard concerns about control over the approximations used to compute the low-energy effective field theory, due to the explicit breaking of supersymmetry by the anti-brane. For the LVS there is also an additional challenge regarding whether an exponentially large volume is consistent with a large warping. We address this issue when we describe LVS models in the first part of this article, and identify the domain of validity for which both are consistent.

A second uplifting proposal obtains de Sitter vacua from D-terms produced by magnetised branes \cite{Dterm,CMQS,kq}. We review this proposal to restate the criteria required to obtain a viable uplifting term from a localised magnetic flux turned on a stack of branes wrapping a large four-cycle. Moreover we also show that fibred Calabi-Yau manifolds open up the possibility
of achieving de Sitter vacua from the interplay of D-terms and string loop corrections to the K\"ahler potential
(and possibly warping).

Our main result is to introduce a novel fully supersymmetric mechanism for obtaining dS solutions in the LVS. This mechanism is based on the presence of the dilaton-dependent non-perturbative effects discussed above, either from gaugino condensation on D3-branes or E(-1)-instantons at singularities.
We show that the corresponding non-perturbative superpotential can give rise to a positive contribution to the scalar potential which differs from the one arising from K\"ahler moduli-dependent non-perturbative effects, as previously considered in KKLT and LVS. These dilaton-dependent non-perturbative effects have never been considered in the context of modulus stabilisation even though non-perturbative superpotentials at a singular locus have been discussed in the past as potential string theory realisations of dynamical supersymmetry breaking and gauge mediation \cite{quiverSUSYbreaking,Romans,Florea,k1}.

We find that combining such a positive contribution with the standard LVS scalar potential gives an uplifting term very similar to the one produced by warped anti D3-brane tension, thereby giving rise to dS vacua in a fully supersymmetric set-up. The fields whose F-terms are responsible
for the realisation of dS vacua are the blow-up modes resolving the singularity of the non-perturbative quiver. Moreover these $S$-dependent non-perturbative effects, when combined to string loop corrections to the K\"ahler potential, can provide a realisation of the LVS also for manifolds with zero or positive Euler number.

This paper is organised as follows. In section \ref{Review} we review the LVS framework and discuss various ways to obtain de Sitter vacua in this setting: uplifting by anti-branes in highly warped throats and D-terms. Readers familiar with the LVS can safely skip this section. Then in section \ref{dSquiver} we analyse the effect of dilaton-dependent non-perturbative effects which provide a novel way to realise de Sitter solutions. This is the main result of the paper. Furthermore, we show that these new effects can also give rise to new LVS for Calabi-Yau
three-folds with positive Euler number due to their interplay with string loop corrections to the K\"ahler potential. In section \ref{Implications} we discuss phenomenological and cosmological implications of this new scenario, and finally conclude in section \ref{Conclusions}. Appendix \ref{AppFieldRedef} is devoted to the study of the effect of loop induced field redefinitions on the form of the scalar potential while in appendix \ref{OnlyNPquiver} we analyse the
subcase with only non-perturbative effects at the singularity.

\section{Review of LVS de Sitter vacua}
\label{Review}

In this section we give a brief review of the basics of the LVS whose main feature is the emergence of a non-supersymmetric AdS minimum at exponentially large value of the internal volume. We will then discuss the main ideas that have been put forward
to obtain de Sitter solutions within the LVS framework working out the conditions under which this can achieved either via anti-branes in highly warped throats or D-terms.

\subsection{LVS in a nutshell}
\label{LVSreview}

The LARGE Volume Scenario corresponds to a concrete mechanism of modulus stabilisation on orientifolded Calabi-Yau compactifications of type IIB string theory. The dilaton and complex structure moduli are fixed by turning on fluxes
of the complex three-form $G_3=F_3+iS H_3$ where $F_3,H_3$ are the Ramond-Ramond and Neveu-Schwarz three-forms and $S=e^{-\phi}+i C_0$ with $\phi$, $C_0$ the dilaton and axion fields of the 10D theory \cite{gkp}. The quantisation of the fluxes gives rise to the discrete landscape of minima.
The K\"ahler moduli are stabilised by a combination of perturbative and non-perturbative effects. The perturbative effects correspond to the leading order term in the expansion in inverse overall volume $\vo$ of the $\alpha'$ corrections which compete with the leading order non-perturbative effect in a blow-up modulus $\tau_s$, leading to the stabilisation at
$\vo\sim e^{a\tau_s}$. The constant $a$ is fixed by the nature of the non-perturbative effect that can be either gaugino condensation on D7-branes or a Euclidean D3-instanton.

It was long thought that this kind of stabilisation was impossible, due to the obstacle identified long ago by Dine and Seiberg \cite{ds} that states that the natural vacuum cannot be analysed in weak coupling since the potential then arises as a series in the coupling constant, which is a field (the dilaton). But then dilaton stabilisation requires a competition among different orders in the weak-coupling expansion, and so cannot be understood within such an expansion.

This objection is overcome in the LVS by realising that in string theory there are actually a number of independent expansions at play, instead of only the one implicitly assumed in \cite{ds}. Besides the string coupling, there are also expansions in inverse powers of all of the various large moduli of the background geometry, of which there can be very many. Even though the minimum identified in the LVS is very generic \cite{ccq2,CMV}, the simplest realisation is for an orientifold of the $\mbb{P}^4_{[1,1,1,6,9]}$ Calabi-Yau having two K\"ahler moduli $\tau_b$, $\tau_s$ and volume $\vo = \tau_b^{3/2} - \tau_s^{3/2}$. The simplest set-up then gives rise to an effective field theory determined by a K\"ahler potential of the type:
\be
K=-2\ln\left(\vo+\frac{\hat\zeta}{2}\right)=-2\ln\left(\tau_b^{3/2}-\tau_s^{3/2}+\frac{\hat\zeta}{2}\right)\,,
\label{KLVS}
\ee
with $\hat\zeta=\zeta/g_s^{3/2}$, and a superpotential:
\be
W=W_0+ A \,e^{-a T_s}\,.
\label{WLVS}
\ee
Here $T_s = \tau_s + i \psi_s$ and $T_b = \tau_b + i \psi_b$ are the complex moduli for $\tau_s$ and $\tau_b$.

After stabilising the axion-like fields, $\psi_s$ and $\psi_b$, the corresponding scalar potential takes the form:
\be
V_{\rm LVS}= c_1  \frac{\sqrt{\tau_s}\,e^{-2a\tau_s}}{\vo} -c_2\frac{ W_0 \tau_s\,e^{-a\tau_s}}{\vo^2} + c_3\frac{W_0^2}{\vo^3}\,,
\label{Vlvs}
\ee
where the $c_i$ are $\mc{O}(1-10)$ constants.
Notice that the three terms of this scalar potential conspire to give a non trivial minimum. The negative sign in the second term comes from the minimisation of the axion component of $T_s$
and drives the minimum to negative values of the vacuum energy. The last term dominates at large values of $\tau_s$ guaranteeing that the potential asymptotes to zero from the positive side (this requires $\zeta>0$). The minimum of this scalar potential is at $\tau_s\sim 1/g_s$ and $\vo \sim W_0\,e^{a\tau_s}$ where the three terms of $V_{\rm LVS}$ are of the same order, giving rise to exponentially large volumes in a natural way. The minimum of this potential corresponds to a non-supersymmetric AdS vacuum.

\subsection{De Sitter LVS}

Several proposals have been put forward to realise a dS solution at exponentially large volume. The most promising so far rely on the inclusion of anti D3-branes at the tip of a warped throat
and D-terms from magnetised D7-branes. Let us briefly review both of these constructions discussing the pros and cons involved.

\subsubsection{Anti-branes and warping}

The simplest proposal to achieve dS space follows KKLT by adding an anti D3-brane at the tip of
a highly warped region somewhere within in the Calabi-Yau space \cite{KKLT}. This contributes a positive term to the scalar potential of the order:
\be
V_{\rm up}= \frac{\nu}{\vo^{\gamma}}\,,
\label{up}
\ee
where $\nu$ can be order unity if $\gamma = 3$, but $\nu$ must be parametrically small if $\gamma\leq 3$. This condition on $\nu$ arises from the requirement that this term can compete with the negative potential generated by the LVS mechanism (in order to uplift it), but not be so large as to dominates at large distances and destabilise the minimum. A small size for the coefficient $\nu$ is plausible if the anti-brane is in a strongly warped region, since then $\nu$ is suppressed by a power of the warp factor, which can be very small.

The analysis of \cite{KKLT} treated the anti-brane in a probe approximation. The effects of the brane were included in the effective field theory by simply considering a potential term equal to the brane tension. The configuration is metastable
with the possibility of decay involving brane flux annihilation \cite{KPV}. There has been much recent work studying the back-reaction of the brane on the geometry \cite{Gr,Gro,Grt,Dym} to develop an understanding of the system which does not rely on the probe approximation.

One finds that the presence of the brane modifies the geometry in the ultraviolet, the associated mode is normalisable supporting the interpretation of the state as a configuration in the same theory.

\subsubsection*{LARGE volume and large warping}

It is very interesting that the two proposals for generating a large hierarchy from extra dimensions --- {\em i.e.} exponentially large volume and large warping --- are realised in a natural way in type IIB string compactifications. Since both can be exponentially large, care must be taken to ensure there are no problems with the validity of the effective field theories which are the main tools for analysis.

We now therefore pause to establish the consistency conditions for having exponentially large volume and large warping at the same time. We identify a common domain of validity for both, and so show that large warping can be possible within the LVS.
We start with a warped ten-dimensional metric of the form:
\be
ds^2= e^{2A} \eta_{\mu\nu} dx^\mu dx^\nu + e^{-2A} g_{mn} dy^mdy^n\,,
\ee
with warp factor $w=e^{-2A}$ and $g_{mn}$ the metric for the Calabi-Yau manifold. The 10D field equations of type IIB supergravity have as solutions $e^{-4A} = e^{-4A_0} + c$ where $c \simeq {\cal O}(\vo^{2/3})$ is an otherwise arbitrary constant and $A_0$ is a profile that depends on the underlying Calabi-Yau geometry. The condition for having a throat that dominates over the volume in part of the geometry is $e^{-4A_0} \gg c$.

But we must also be sure that the warping is not so strong as to invalidate the 4D effective field theory at energies below the warped Kaluza-Klein scale. This requires in particular that the moduli masses must be much smaller than the warped KK scale. Requiring this to be true in a region where warping dominates large-volume in the geometry therefore leads to the twin conditions \cite{warping}:
\be
e^{-4A_0}>\vo^{2/3} > e^{-A_0}\,.
\label{constraint}
\ee
Here $A_0$ is considered at its maximum value (tip of the throat). The first of these is the condition that the effective warp factor is dominated by $A_0$ rather than $c$:  $e^{-4A} \simeq e^{-4A_0}$ with $c\sim \vo^{2/3}$. The second condition is the condition that modulus masses are not higher than the warped KK scale, which requires $e^{A_0} > \vo^{-2/3}$ \cite{warping}.

For the present purposes, what is important is that we must check if any candidate warped uplifting term --- as in KKLT and LVS scenarios --- is consistent with these constraints. That is, uplifting potentials usually have the form:
\be
V_{\rm up} \sim \frac{e^{4A_0}}{\vo^{\alpha-2/3}}\,,
\label{uplift1}
\ee
where the unwarped contribution would go like $1/\vo^{\alpha}$ and for an anti-D3 brane $\alpha=2$. In the LVS case, for tuning the cosmological constant, the uplifting term (\ref{uplift1}) has to be comparable in size to the rest of the potential (\ref{Vlvs}), and so we need:
\be
\frac{e^{4 A_0}}{\langle \vo \rangle^{\alpha -2/3} } \sim \frac{1}{\langle \vo \rangle^3}
\quad \Leftrightarrow \quad
e^{- A_0} \sim \langle \vo \rangle^{11/12-\alpha/4}\,.
\label{eap}
\ee
This satisfies the above constraint (\ref{constraint}) for $\alpha$ in the range:
\be
1<\alpha<3\,,
\ee
which is satisfied in particular by anti-D3 branes, for which $\alpha=2$. Therefore it can be consistent to realise de Sitter uplifting via anti-D3 branes at the tip of a warped throat in the LARGE Volume Scenario.

\subsubsection{D-terms from magnetised branes revisited}

Another interesting mechanism to achieve dS vacua relies on D-terms from magnetised D7-branes since they give rise to a positive contribution to the scalar potential of the form (\ref{up}). Concrete examples have been worked out in \cite{Dterm,CMQS,CGJR}. Notice that in KKLT scenarios D-terms are not appropriate for uplifting since the AdS minimum is supersymmetric and vanishing F-terms imply vanishing D-terms also. On the other hand, the situation is much better in LVS since the AdS minimum is non supersymmetric and therefore D-terms can be non-zero.

The D-term potential associated to the diagonal $U(1)$ factor
of a stack of D7-branes wrapping the Calabi-Yau divisor $D_i$,
looks like \cite{HKLVZ}:
\be
V_D = \frac{g_i^2}{2} \left( \sum_j c_{ij} \hat{K}_j\varphi_j -\xi_i\right)^2\,,
\label{VD}
\ee
where $\hat{K}$ is the matter K\"ahler potential.
In the previous expression the gauge coupling is given by $g_i^{-2}= {\rm Vol}(D_i) /(4\pi) =\tau_i/(4\pi)$,
$\varphi_j$ are matter fields (open string modes) with $U(1)$ charges $c_{ij}$, while $\xi_i$ is
the Fayet-Iliopoulos term. This term is generated by turning on a magnetic flux $\mc{F}_i=\tilde{f}_i^k \hat{D}_k$
on the D7-brane stack wrapping $D_i$:
\be
\xi_i=\frac{1}{4\pi\vo}\int_X \hat{D}_i \wedge J\wedge \mc{F}_i
=-\frac{q_{ij}}{4\pi}\,\frac{\partial K}{\partial T_j}\,,
\label{FI}
\ee
where $q_{ij}$ are the $U(1)$ charges of the K\"ahler moduli which can be expressed in terms of
the triple intersection numbers $k_{ijk}$ of the Calabi-Yau as:
\be
q_{ij}= \tilde{f}^k k_{ijk}\,.
\label{triple}
\ee
Ref. \cite{CMQS} studied the interplay between F- and D-terms in the LVS in order to obtain dS vacua, finding that an attractive candidate can be found when the following three criteria are satisfied:
\begin{itemize}
\item The magnetised D7-brane stack wraps the large four-cycle;
\item The K\"ahler modulus $T_b$ (described in section \ref{LVSreview}) is charged under the $U(1)$;
\item One of the open string modes, stretching between the D7-brane with magnetic flux and its orientifold image, is tachyonic at the origin of field space. Such a situation is fairly generic and can be shown to arise as a result of relationships between the charges of various open string fields connected to anomaly cancellation conditions.
\end{itemize}
Under such a situation, ref. \cite{CMQS} found that, upon minimisation, the combined F- and D-term potential yields a contribution which scales as $\vo^{-8/3}$ which is good since $1<8/3<3$.

However, such a contribution can uplift the AdS vacuum to dS
in a controlled fashion only if the magnetic flux on the brane is localised in a warped region. This requires that a two-cycle of the large four-cycle be deep in a warped throat which is not impossible but not straightforward to achieve, as we now see.

To see why it can be hard to have localised magnetic flux which can generate the uplift term of \cite{CMQS}, note first that
we require that the brane wraps the large cycle and also that the K\"ahler modulus $T_b$ is charged under the $U(1)$, i.e:
\be
q_{bb} = \tilde{f}^i k_{i b b} \neq 0\,.
\label{bcharge}
\ee
In order for the magnetic flux to be localised, no flux can thread the large cycle since such a contribution would
have support in the entire Calabi-Yau, and so it is difficult to conceive that the
associated energy can be lowered by warping. Thus we would like to set $\tilde{f}^b =0$,
and have non trivial flux threading a small cycle, i.e $\tilde{f}^s \neq 0$. Then the condition (\ref{bcharge}) implies that:
\be
k_{sbb} \neq 0\,.
\ee
However ref. \cite{CKM} pointed out that in order
to find a LVS the small four-cycle supporting the non-perturbative effects has to be a `diagonal' del Pezzo divisor which is characterised by the fact that there exists a basis of toric divisors where the only non-vanishing intersection number is $k_{sss}\neq 0$. Hence this four-cycle enters the overall volume in a completely diagonal
way, showing that it can be shrunk to zero size without affecting the Calabi-Yau geometry. Therefore in this case the condition $k_{sbb} \neq 0$ cannot be realised.

However this condition might still be realised if the number of K\"ahler moduli is greater than two and there are other small four-cycles which do not support non-perturbative effects
as in \cite{CMV}. These additional small four-cycles support a GUT- or MSSM-like visible sector and, in order to avoid the shrinking induced by D-terms, have to be `non-diagonal' rigid
but not del Pezzo divisors, according to the classification of \cite{CKM}. Then one can indeed satisfy the condition $k_{sbb} \neq 0$ \cite{CMV}.

\subsubsection*{D-term uplift for fibred Calabi-Yau manifolds}

An alternative way to achieve dS vacua via D-terms without having to deal with localised magnetic flux,
has been proposed in \cite{CGJR} for Calabi-Yau three-folds where the overall volume is controlled by two
four-cycles $\tau_1$ and $\tau_2$:
\be
\vo=\sqrt{\tau_1}\tau_2-\tau_s\,.
\ee
The sum of F- and D-term potential coming from a magnetised D7-brane stack wrapping $\tau_1$ is:
\be
V=V_D + V_F = \frac{\pi}{\tau_1} \left(
q_{\phi} |\phi_c|^2-\xi_1 \right)^2+k \frac{W_0^2}{\vo^2}\,|\phi_c|^2+V_{\rm LVS}\,,\quad\text{with}\quad\xi_1=p\frac{\sqrt{\tau_1}}{\vo}\,,
\label{NewVtotale}
\ee
where $\phi_c$ is a canonically normalised matter field while $k$, $q_{\phi}$ and $p$ are all $\mc{O}(1)$ numbers. The LVS potential is as described above, and depends only on the moduli $\tau_s$ and $\vo$.

The minimum for $\phi_c$ is at:
\be
\langle|\phi_c|^2\rangle= \frac{\xi_1}{q_{\phi}}-\frac{k W_0^2 \,\tau_1}{2\pi q_{\phi}^2 \vo^2}\simeq \frac{\xi_1}{q_{\phi}}\,,
\ee
where the approximate equality uses $\xi_1 \simeq \sqrt{\tau_1}/\vo$ and $\vo \gg 1$. Eq.~(\ref{NewVtotale}) then reduces to:
\be
V = \lambda W_0^2 \frac{\sqrt{\tau_1}}{\vo^3}+V_{\rm LVS}\,,\quad\text{with}\quad\lambda=\frac{k p}{q_{\phi}}\,.
\label{NewVsu}
\ee

It is now a problem that the potential $V_{\rm LVS}$ (\ref{Vlvs}) does not depend on $\tau_1$, because the modulus $\tau_1$ cannot be fixed. However we know that another term in $V$ depends on $\tau_1$ once string-loop corrections to $K$ are included, coming from loops of open strings living on the D7-branes wrapped around $\tau_1$ \cite{ccq1}. Including also this $\tau_1$-dependent part of the scalar potential we find:
\be
V =\left(\lambda\frac{\sqrt{\tau_1}}{\vo}+\frac{g_s^2 c_{\rm loop}^2}{\tau_1^2}\right)\frac{W_0^2}{\vo^2}\,,
\label{Vcombined}
\ee
where $c_{\rm loop}$ is a loop coefficient that depends on the complex structure moduli. The potential (\ref{Vcombined}) has a minimum for $\tau_1$ at:
\be
\tau_1 =\left(4 g_s^2 c_{\rm loop}^2/\lambda\right)^{2/5}\vo^{2/5}\,,
\label{tau1VEV}
\ee
which substituting back in (\ref{Vcombined}) gives:
\be
V =\mu\frac{W_0^2}{\vo^{14/5}}+V_{\rm LVS}\,,
\quad\text{with}\quad\mu=5\left(\frac{\lambda}{4}\right)^{4/5}\left(g_s c_{\rm loop}\right)^{2/5}.
\label{A}
\ee

We see that in order to get a Minkowski vacuum, the coefficient $c_{\rm loop}$ must be tuned such that \cite{CGJR}:
\be
\mu \sim \mc{O}(1) \,g_s^{-1/2}\langle\vo \rangle^{-1/5}\qquad\Leftrightarrow\qquad c_{\rm loop}\sim \mc{O}(1)\, g_s^{-9/4}\langle\vo \rangle^{-1/2}\,.
\ee
Substituting this tuning back in (\ref{tau1VEV}) we realise that:
\be
\langle\tau_1\rangle \sim g_s^{-1}\sim \mc{O}(10)\,.
\ee
Notice that the tuning of the loop coefficient includes the value of the volume
at the minimum which may not be natural for very large volumes. However there is always
the possibility that the corresponding cycle lies within a warped region for which the coefficient $\lambda$
is warped and volume dependent and therefore the tuning is similar to the
antibrane uplifting in KKLT.

We finally mention that another mechanism to achieve dS vacua that has been proposed, relies on F-terms from a hidden sector with metastable susy breaking \cite{ss}. However this mechanism has not yet been implemented within the LVS. It would be very interesting to have a concrete realisation
of this mechanism in the LVS.

\section{De Sitter vacua from non-perturbative effects}
\label{dSquiver}

Non-perturbative effects play a major r\^ole in stabilising the K\"ahler moduli $T_i$ of type IIB Calabi-Yau flux compactifications. In fact, the starting point of KKLT and LVS scenarios is the inclusion of $T_i$-dependent
non-perturbative corrections to the superpotential of the form:
\be
W_{\rm np} =\sum_i A_i\, e^{-\,a_i \, f_i}\,,
\label{WnpinT}
\ee
where the prefactors $A_i$ are in general functions of the complex structure moduli $U$
and the open string moduli (D7-deformation and D3-position moduli), the $a_i$ are constants and
the $f_i$ are the corresponding gauge kinetic functions which are given by:
\be
f_i = T_i + h_i(\mc{F}) S\,,
\label{gkfinT}
\ee
where $T_i$ is the K\"ahler modulus controlling the size of the divisor $\Sigma_i$
wrapped by either a stack of D7-branes or a Euclidean D3-instanton,
while $h$ is a function of the world-volume magnetic flux $\mc{F}$.
Let us briefly describe the two cases for $\mc{F}=0$ and $\mc{F}\neq 0$ which give rise to
two physically different situations.

\begin{itemize}
\item {\it Vanishing magnetic flux: $\mc{F}=0$}. In this case $h_i=0$ and
the non-perturbative effects have three different microscopic realisations:
\begin{itemize}
\item {\it Divisor $\Sigma_i$ wrapped just by a stack of D7-branes}:
In this case the superpotential (\ref{WnpinT}) is generated by gaugino condensation.
The simplest realisation of a pure $\mc{N}=1$ super Yang-Mills theory that undergoes
gaugino condensation is via a stack of D7-branes wrapping an orientifold-invariant four-cycle without any
magnetic flux on the D7-branes. The constants $a_i$ are given by $a_i=6\pi/\beta_i$ where $\beta_i$ are
the one-loop beta function coefficients of the condensing gauge theories. For example,
in the case of a pure $SU(N)$ theory, $\beta_i= 3 N$ while the in case of a pure $Sp(2N)$ theory,
$\beta_i = 3 (N+1)$.

\item {\it Divisor $\Sigma_i$ wrapped just by a Euclidean D3-brane instanton}:
In this case the superpotential (\ref{WnpinT}) is generated by a so-called `stringy' instanton.
The simplest realisation which yields the right fermionic zero mode structure,
involves a rigid four-cycle which is transversally invariant under the orientifolds and
a vanishing magnetic flux on the instanton. In this case the constants $a_i$ are given by $a_i=2\pi$.

\item {\it Divisor $\Sigma_i$ wrapped by both a stack of D7-branes and a Euclidean D3-brane instanton}:
In this case the main contribution to the superpotential (\ref{WnpinT}) comes from gaugino condensation since
the contribution due to E3-instantons is more suppressed due to the different behaviour of the constants $a_i$.
\end{itemize}

\item {\it Non-vanishing magnetic flux: $\mc{F}\neq 0$}. In this case $h_i\neq 0$ and
the non-perturbative effects can have again three different microscopic realisations:

\begin{itemize}
\item {\it Divisor $\Sigma_i$ wrapped just by a stack of D7-branes}:
In the presence of a non-vanishing magnetic flux $\mc{F}$ on the world-volume of a D7-brane stack,
chiral matter gets generated, and so the theory undergoes gaugino condensation only for particular configurations.
For examples an $SU(N_c)$ theory with $N_f$ flavours, undergoes gaugino condensation only for $N_f<N_c-1$.
Moreover, in this case the K\"ahler moduli get charged under the anomalous $U(1)$, and so the prefactors
$A_i$ have to depend also on charged matter fields whose $U(1)$ transformation has to compensate the shift transformation of the $T_i$ fields
in order to render the superpotential (\ref{WnpinT}) gauge invariant.

\item {\it Divisor $\Sigma_i$ wrapped just by a Euclidean D3-brane instanton}: Given that $\mc{F}$ is anti-invariant
under the orientifold and in order to have a non-vanishing contribution to the superpotential the instanton configuration
has to be invariant under the orientifold, we can have $\mc{F}\neq 0$ only if the magnetic flux is purely odd,
i.e. $\mc{F} \in H^{1,1}_-(\Sigma_i)$ \cite{GKPW}. This is possible only for orientifold projections such that $h_{1,1}^-\neq 0$.
In this case the gauge kinetic functions acquire a dependence on the $G$-moduli associated with the $h_{1,1}^-$ odd four-cycles,
$f_i = T_i + h_{ij} (\mc{F}) G_j$.

\item {\it Divisor $\Sigma_i$ wrapped by both a stack of D7-branes and a Euclidean D3-brane instanton}:
In this case the superpotential (\ref{WnpinT}) is generated by a so-called `gauge' instanton corresponding to the case $N_f=N_c-1$
and the prefactors $A_i$ have again to depend on charged matter field in order to guarantee the gauge invariance of the superpotential.
\end{itemize}
\end{itemize}

Scenarios with vanishing magnetic flux are simpler for modulus stabilisation.
The case with $\mc{F}\neq 0$ includes more general expressions but also constraints.
In particular, it complicates the stabilisation of the four-cycle supporting the Standard Model (SM)
by these non-perturbative effects since requiring non-vanishing coefficients $A_i$
would generically break the SM symmetry at high energies by the non-vanishing values of the charged matter fields \cite{Tension}.
In those cases D-terms or perturbative, rather than non-perturbative, effects are preferred to fix the SM cycle \cite{CMV}.

The non-perturbative effects described above are the only ones considered so far for modulus stabilisation in type IIB flux
compactifications. However, there are further dilaton-dependent non-perturbative effects generated by
either gaugino condensation on spacetime-filling D3-branes or Euclidean D(-1)-brane (E(-1)) instantons.
Given that the associated branes do not wrap any internal cycle,
the corresponding gauge kinetic function is determined by the dilaton field $S$:
\be
f_i = S +h_i \,Q_i\,,
\label{gkfinS}
\ee
where, in analogy with (\ref{gkfinT}), we have included a shift proportional to $h$. As we have seen,
in the case of D7-branes and E3-instantons, $h$ is non-zero only in the presence of a magnetic flux.
In the present case, this is equivalent to place the D3-branes and/or the E(-1)-instantons at a singularity.
In fact, as a magnetic flux is responsible for the emergence of chirality in the geometric case,
here chiral matter gets generated by the presence of the singularity. Hence the $h_i$ in (\ref{gkfinS})
are constants proportional to the $U(1)$-charge of the blow-up modes $Q_i$ resolving the singularity.

In parallel with the previous discussion for the geometric case, we have two cases for $h_i=0$
(branes at smooth points and no chiral matter)
and $h_i\neq 0$ (branes at singularities and chiral matter).
For each case we have again three different microscopic realisations corresponding to:
\begin{itemize}
\item {\it Stack of spacetime-filling D3-branes}:
In this case the superpotential (\ref{WnpinT}) is generated by gaugino condensation.
The simplest realisation of a pure $\mc{N}=1$ super Yang-Mills theory that undergoes
gaugino condensation is via a stack of spacetime-filling D3-branes at a smooth point on top of the O-plane.
If the D3-branes are located at a singularity chiral matter gets generated,
and so the theory undergoes gaugino condensation only for particular configurations.
Moreover, in this case the blow-up modes $Q_i$ get charged under the anomalous $U(1)$, and so the prefactors
$A_i$ have to depend also on charged matter fields whose $U(1)$ transformation has to compensate the shift of the $Q_i$ fields
in order to render the superpotential gauge invariant.
We finally point out that the location of the D3-branes can be determined by minimising the potential
for the D3-position moduli $\zeta_i$ which is generated by different effects ($\zeta_i$-dependence of the K\"ahler potential and
the prefactors $A_i$ plus D-terms) \cite{malda}.

\item {\it Euclidean D(-1)-brane instanton}:
In this case the superpotential (\ref{WnpinT}) is generated by a so-called `stringy' instanton.

\item {\it Euclidean D(-1)-brane instanton on top of a stack of spacetime-filling D3-branes}:
When the branes are located at smooth points, the main contribution to the superpotential (\ref{WnpinT}) comes from gaugino condensation since
the contribution due to E(-1)-instantons is more suppressed due to the different behaviour of the constants $a_i$.
On the other hand, for branes at singularities, the superpotential (\ref{WnpinT}) is generated by a so-called `gauge' instanton
corresponding to the case $N_f=N_c-1$ and the prefactors $A_i$ have again to depend on charged matter fields in order to guarantee the gauge invariance of the superpotential.
\end{itemize}

In this section we will explore a combination of both classes of non-perturbative effects
within the LVS framework.
We shall focus, without loss of generality,
on the case with a single K\"ahler moduli-dependent and a single dilaton-dependent contribution
to the non-perturbative superpotential:
\be
W_{\rm np} = A\, e^{-\,a \, T } + B\, e^{-\, b \, \left(S+ h Q\right)}\,,
\label{TotalWnp}
\ee
where we will consider just one blow-up mode $Q$.
We will find that the dilaton-dependent non-perturbative effect proportional to $B$
can give rise to de Sitter vacua only for $h\neq0$. In fact, in the case $h=0$, the new $S$-dependent
non-perturbative effect can be included in a redefinition of $W_0$ $\to$ $W_0'\equiv W_0 + B\,e^{-b\,S}$,
showing that this term cannot give rise to any new interesting dynamics.

We shall find that for $h\neq 0$ it is the F-term of the blow-up mode $Q$ which is responsible for uplifting
since it gives rise to
a positive definite term in the scalar potential of the form $e^{-2\,b {\rm Re}(S)}/\vo$.
Given that ${\rm Re}(S)$ is fixed by ratios of flux integers,
this term has a similar effect as the standard warped anti D3-brane contribution
with the advantage that it comes from a manifestly supersymmetric effective action.

As is well known, the LVS applies for Calabi-Yau compactifications with negative Euler number ($h^{1,2}>h^{1,1}$).
However in this section we shall also present another interesting implication of adding an $S$-dependent $W_{\rm np}$:
the possibility to obtain exponentially large volumes and de Sitter space also for positive and vanishing Euler numbers.
This requires the introduction of string loop contributions to the effective action which are less understood.
In appendix \ref{OnlyNPquiver} we will discuss briefly a simplified case with non-perturbative effects only at the singularity regime.

\subsection{De Sitter LVS from D3/E(-1) non-perturbative effects at singularities}
\label{NPsing}

Let us study how the LVS stabilisation mechanism gets modified by the inclusion of
dilaton-dependent non-perturbative effects. We shall analyse the case for $h\neq 0$ corresponding to D3-branes and/or E(-1)-instantons at singularities
since, as we have pointed out above, the case for $h=0$ does not give rise to any new result.

As reviewed in section \ref{LVSreview}, the standard LVS involves at least two K\"ahler moduli,
$T_b=\tau_b+{\rm i}\,\psi_b$ and $T_s=\tau_s + {\rm i}\,\psi_s$, which can be fixed by the interplay of the leading order $\alpha'$
correction to the K\"ahler potential and a single $T_s$-dependent non-perturbative superpotential.
Hence we shall extend the superpotential (\ref{WLVS}) as in (\ref{TotalWnp}) obtaining:
\be
W=W_0(S)+ A\, e^{-a T_s}+B \,e^{-b \left(S+h Q\right)}\,,
\label{Wnon-pert}
\ee
where $Q=\rho + {\rm i}\,\psi_{\rho}$ is the modulus that blows up the singularity ($\rho=0$ corresponds to a collapsed four-cycle) and $h\neq 0$. \footnote{This extra sector can also be considered for the KKLT scenario for which there will be a non-perturbative superpotential for the overall K\"ahler modulus $T$ and also for the blow-up mode $Q$. However since the original minimum was supersymmetric, $D_TW=0$, as long as there is a solution to $D_QW=0$ the vacuum remains supersymmetric AdS; although such a solution is likely to be away from the singular locus.}
The K\"ahler potential (\ref{KLVS}) of the effective supersymmetric field theory has to be supplemented with the term for the blow-up mode $\rho$,
and so it takes the more general form (writing $s = {\rm Re}(S)$):
\be
K=-\ln(2s)-2\ln \left(\vo+\frac{\zeta \,s^{3/2}}{2}\right)+\alpha \frac{\rho^2}{\vo}\,,
\label{KahlerPot}
\ee
with:
\be
\vo= \tau_b^{3/2}-\tau_s^{3/2}\,.
\ee
Notice that, even if the dilaton is fixed at leading semi-classical order by imposing $D_S W=0$,
both in the superpotential (\ref{Wnon-pert}) and in the K\"ahler potential (\ref{KahlerPot})
we kept the dilaton dependence manifest in order to have control over higher order contributions to the scalar potential
which could play an important r\^ole for achieving dS vacua.

In the large volume limit $\tau_b\gg \tau_s$, the K\"ahler potential (\ref{KahlerPot}) can be expanded as:
\be
K\simeq -\ln(2s)-3\ln \tau_b +2 \,\epsilon_{\tau_s}  - \frac{\zeta \,s^{3/2}}{\tau_b^{3/2}}\left(1+\epsilon_{\tau_s} \right)
+\alpha \frac{\rho^2}{\tau_b^{3/2}}\left(1+\epsilon_{\tau_s} \right)\,,
\label{KPot}
\ee
where
\be
\epsilon_{\tau_s} \equiv \left(\frac{\tau_s}{\tau_b}\right)^{3/2}\sim \mc{O}(\vo^{-1})\ll 1\,.
\ee

We point that the moduli $T_b$ and $T_s$ are both
larger than the string scale, while $Q$ is at the singular locus. It is worth stressing that
the effective field theory can be trusted not just in the case when the moduli are in the geometric regime,
but also in the regime close to a singularity thanks to the detailed description
of strings at orbifold singularities. The field $Q$ can indeed be shown to be fixed
at the singular locus as a result of
D-term stabilisation. If we turn on a flux on a two-cycle inside the four-cycle whose volume is given by the blow-up mode $\rho$,
the D-term potential takes the same form as in (\ref{VD}):
\be
V_D = \frac{g^2}{2} \left( \sum_i c_i \hat{K}_i\varphi_i -\xi\right)^2\,,
\ee
where, according to (\ref{FI}), the FI-term $\xi$ becomes:
\be
\xi=-\frac{q_{\rho}}{4\pi} \frac{\partial K}{\partial Q}=-\frac{\alpha q_{\rho}}{4\pi} \frac{\rho}{\vo}\,.
\ee
Here $q_{\rho}$ is the $U(1)$ charge of the blow-up mode $\rho$
that depends on the magnetic flux and its triple self intersection number (see (\ref{triple})).

In this section we shall proceed assuming that the D-term potential can be minimised
with non-zero VEVs for some charged matter fields such that $\langle\sum_i c_i \hat{K}_i\varphi_i\rangle=0$
together with a non-vanishing prefactor $B$
of the non-perturbative superpotential ($B\neq 0$). \footnote{The situation is similar to the tension between moduli
stabilisation by non-perturbative effects and chirality discussed in \cite{Tension};
although in our case this results into a much weaker constraint as the open string fields belong to a hidden sector,
and so can get non-zero VEVs.}
The details of such settings will be
elaborated in section \ref{generation}. Thus the D-term scalar potential reduces to:
\be
V_D= \frac{g^2}{2}\,\xi^2=\frac{g^2}{2}\left(\frac{\alpha q_{\rho}}{4\pi}\right)^2 \frac{\rho^2}{\vo^2}
\simeq \frac{(\alpha q_{\rho})^2}{8\pi}
\frac{\rho^2}{(s+h\rho) \tau_b^3}\left(1+2\epsilon_{\tau_s}\right),
\ee
which uses $4\pi g^{-2} = s + h\rho$ and $\vo^{-2} \simeq \tau_b^{-3}(1 + 2 \epsilon_{\tau_s})$.

We now turn to the computation of the F-term scalar potential:
\be
V_F =e^K\left(K^{I\bar{J}}D_I W D\overline{W}_{\bar{J}}-3|W|^2\right),
\ee
where the K\"ahler covariant derivative is $D_I W= W_I+W K_I$.
The K\"ahler metric is a symmetric matrix which reads (with the leading order correction in a large volume expansion):
\be
K_{I\bar{J}}=\left(
\begin{array}{cccc}
 \frac{1}{4 s^2}\left(1-\frac{3\epsilon_s}{4} \right) &
 \frac{9  \sqrt{s}\, \zeta }{16 \tau_b^{5/2}}(1+2\epsilon_{\tau_s}) & -\frac{9 \sqrt{s\,\tau_s} \,\zeta
   }{16 \tau_b^3} & 0 \\
 K_{1\bar{2}} & \frac{3}{4
   \tau_b^2}\left(1+\frac{ 10 \epsilon_{\tau_s} + 5\epsilon_{\rho}-5\epsilon_s}{4} \right) & -\frac{9  \sqrt{\tau_s}}{8 \tau_b^{5/2}}(1+\epsilon_{\rho}-\epsilon_s) & -\frac{3 \alpha  \, \rho}{4 \tau_b^{5/2}}(1+2\epsilon_{\tau_s}) \\
K_{1\bar{3}} & K_{2\bar{3}} & \frac{3 }{8
   \tau_b^{3/2} \sqrt{\tau_s}} \left(1+\frac{\epsilon_{\rho}-\epsilon_s}{2} \right)& \frac{3 \alpha \,\rho \sqrt{\tau_s}}{4 \tau_b^3} \\
 K_{1\bar{4}} & K_{2\bar{4}} & K_{3\bar{4}} & \frac{\alpha }{2 \tau_b^{3/2}}(1+\epsilon_{\tau_s})
\end{array}
\right), \nn
\ee
where:
\be
\epsilon_{\rho}\equiv \frac{\alpha \, \rho^2}{\tau_b^{3/2}}\sim \mc{O}(\vo^{-1})\ll 1
\qquad\text{and}\qquad\epsilon_s\equiv \zeta\left(\frac{s}{\tau_b}\right)^{3/2}\sim \mc{O}(\vo^{-1})\ll 1\,.
\ee
The inverse K\"ahler metric with the leading order correction in a large volume expansion reads:
\be
\left(
\begin{array}{cccc}
 4 s^2 \left(1+\frac{3\epsilon_s}{4} \right) &
 -\frac{3 s^{5/2}\, \zeta}{\sqrt{\tau_b}}\left(1+\frac{4\epsilon_{\tau_s} + 8\epsilon_s + \epsilon_{\rho}}{4} \right) &
 -\frac{3 s^{5/2} \, \zeta \tau_s }{\tau_b^{3/2}}\left(1+ \frac{12 \epsilon_{\tau_s} + 8 \epsilon_s + \epsilon_{\rho}}{4} \right) &
 -\frac{9 s^{5/2} \,\zeta \rho}{2 \tau_b^{3/2}}\left(1+\frac{4\epsilon_{\tau_s}+8\epsilon_s+ \epsilon_{\rho}}{4}\right) \\
 K^{1\bar{2}} &
 \frac{4}{3} \tau_b^2 \left(1+\frac{8\epsilon_{\tau_s}+ 5\epsilon_s+\epsilon_{\rho}}{4} \right) &
 4 \tau_b \tau_s \left(1+\frac{8\epsilon_{\tau_s}+ 3\epsilon_s-\epsilon_{\rho}}{4} \right) &
 2 \,\rho \tau_b \left(1+\frac{5\epsilon_s+\epsilon_{\rho}}{4} \right) \\
 K^{1\bar{3}} & K^{2\bar{3}}  &
 \frac 83 \tau_b^{3/2} \sqrt{\tau_s} \left(1+\frac{9\epsilon_{\tau_s} + \epsilon_s - \epsilon_{\rho}}{2} \right) &
 2 \,\rho \tau_s\left(1+\frac{8\epsilon_{\tau_s}+ 5\epsilon_s + \epsilon_{\rho}}{4} \right) \\
 K^{1\bar{4}} & K^{2\bar{4}} & K^{3\bar{4}} & \frac{2\tau_b^{3/2}}{\alpha}\left(1+\frac{3\epsilon_{\rho} - 2\epsilon_{\tau_s}}{2}\right)
\end{array}
\right). \nn
\ee
Therefore the F-term scalar potential is given by:
\bea
V_F&=&\frac{(1+\epsilon_{\rho}) (1-\epsilon_s) (1+2\,\epsilon_{\tau_s})}{2 s \tau_b^3} \left\{ K^{S\bar{S}} D_{\bar{S}}\overline{W} D_S W
-a A K^{S \bar{T_s}} \left(D_S W e^{{\rm i} a \psi_s}+D_{\bar{S}} \overline{W} e^{-{\rm i} a \psi_s}\right) e^{-a \tau_s}
\right. \nn \\
&& \left. -h b B K^{S\bar{Q}} \left(D_S W e^{+ {\rm i }b (C_0 + h \psi_{\rho})}
+D_{\bar{S}} \overline{W} e^{- {\rm i} b (C_0 + h \psi_{\rho})}\right)
e^{-b (s+ h \rho)}
+ (\overline{W} D_S W+W D_{\bar{S}}\overline{W}) K^{Si} K_i \right. \nn \\
&& \left. + (a A)^2 K^{T_s \bar{T_s}} e^{-2 a \tau_s}-2 a A W \cos\left(a \psi_s\right)
K^{T_s i} K_i\, e^{-a \tau_s} \right. \nn \\
&& \left. + (h b B)^2 K^{Q \bar{Q}}  e^{-2 b (s+ h \rho)}
-2 h b B W \cos\left[ b \left(C_0 +h \psi_{\rho}\right)\right]
K^{Qi} K_i \,e^{-b (s +h \rho)} \right. \nn \\
&& \left. + 2 h a A b B K^{T_S \bar{Q}} \cos\left[a \psi_s + b \left(C_0 + h \psi_{\rho}\right)\right]
e^{- a \tau_s - b (s + h \rho)}
+W^2 \left(K^{ij} K_i K_j-3\right)\right\}.
\eea
Substituting the values of the elements of the inverse K\"ahler metric we obtain:
\bea
V_F&=&\frac{(1+\epsilon_{\rho}) (1-\epsilon_s) (1+2\,\epsilon_{\tau_s})}{2 s \tau_b^3} \left\{ 4 s^2 D_{\bar{S}}\overline{W} D_S W
\left(1+\frac{3\epsilon_s}{4} \right) \right. \nn \\
&& \left. +3 a A \zeta\,s^{5/2} \,\tau_s \left(D_S W e^{{\rm i} a \psi_s}+D_{\bar{S}} \overline{W} e^{-{\rm i} a \psi_s}\right)
\frac{e^{-a \tau_s}}{\tau_b^{3/2}}\left(1+ \frac{12 \epsilon_{\tau_s} + 8 \epsilon_s +\epsilon_{\rho}}{4} \right)
\right. \nn \\
&& \left. + \frac 92 h b B \zeta\,s^{5/2} \, \rho \left(D_S W e^{+ {\rm i }b (C_0 + h \psi_{\rho})}
+D_{\bar{S}} \overline{W} e^{- {\rm i} b (C_0 + h \psi_{\rho})}\right) \frac{e^{-b (s+ h \rho)}}{\tau_b^{3/2}}
\left(1+\frac{4\epsilon_{\tau_s}+8\epsilon_s+\epsilon_{\rho}}{4}\right)\right. \nn \\
&& \left. + \frac 92 \zeta (\overline{W} D_S W+ W D_{\bar{S}}\overline{W}) \frac{s^{5/2}}{\tau_b^{3/2}}
\left(1+ \frac{4\epsilon_{\tau_s}+6 \epsilon_s-\epsilon_{\rho}}{4}\right)
+W^2 \left(3+\frac{3 \epsilon_s- \epsilon_{\rho}}{4}-3\right)\right. \nn \\
   && \left. + \frac 83  (a A)^2  \sqrt{\tau_s}\, \tau_b^{3/2}\,e^{-2 a \tau_s}
   \left(1+\frac{9\epsilon_{\tau_s} + \epsilon_s - \epsilon_{\rho}}{2} \right)
   + 4 a A W \cos\left(a \psi_s\right) \tau_s\, e^{-a \tau_s}
   \left(1+\frac{3\epsilon_s-\epsilon_{\rho}}{4}\right)   \right. \nn \\
   && \left. + \frac{2}{\alpha}(h b B)^2 \tau_b^{3/2}  e^{-2 b (s+ h \rho)}
   \left(1+\frac{3 \epsilon_{\rho} - 2\epsilon_{\tau_s}}{2}\right)
   +2 h b B W \cos\left[ b \left(C_0 +h \psi_{\rho}\right)\right] \rho \,e^{-b (s +h \rho)}
   \left(1+\frac{9\epsilon_s - 3 \epsilon_{\rho}}{4}\right) \right. \nn \\
   && \left. + 4 h a A b B \,\rho \tau_s \cos\left[a \psi_s + b \left(C_0 + h \psi_{\rho}\right)\right]
   e^{- a \tau_s - b (s + h \rho)}\left(1+\frac{8\epsilon_{\tau_s}+5 \epsilon_s +\epsilon_{\rho}}{4} \right)
   \right\}. \nn
\eea
Notice that the K\"ahler covariant derivative with respect to the dilaton scales as:
\be
D_S W = \underset{\mc{O}(1)}{\underbrace{D_S W_0|_{\zeta=0}}}
- \underset{\mc{O}(\vo^{-1})}{\underbrace{\left[\frac 34 \frac{\epsilon_s}{s} W_0(S)+\left(b+\frac{1}{2\,s}\right) B e^{-b(S+hQ)}\right]}}
-\underset{\mc{O}(\vo^{-2})}{\underbrace{\frac 34 \frac{\epsilon_s}{s} B e^{-b(S+hQ)}}}\,,
\label{DSW}
\ee
where:
\be
D_S W_0|_{\zeta=0}=\left(\frac{\partial_s W_0}{2}-\frac{W_0}{2s}\right)\,.
\ee
The volume scaling of (\ref{DSW}) follows from the fact that we shall be interested in studying the
scalar potential in the large volume limit $a \tau_s\sim b s\sim \ln \vo\gg 1$.

The total scalar potential $V=V_D+V_F$ receives contributions at different orders in a large volume expansion.
It is therefore convenient to study its behaviour order by order in $1/\vo$ writing:
\be
V= V_{\mc{O}(\vo^{-2})}+V_{\mc{O}(\vo^{-3})}+V_{\mc{O}(\vo^{-4})}+\dots \,,
\ee
where:
\be
V_{\mc{O}(\vo^{-2})}=\frac{1}{\tau_b^3} \left[ 2 s | D_S W_0|_{\zeta=0}|^2
+\frac{(\alpha q_{\rho})^2}{8\pi}\frac{\rho^2}{\left(s+h\rho\right)}\right].
\ee
and:
\bea
V_{\mc{O}(\vo^{-3})}&=&\frac{1}{\tau_b^3}
\left\{2 s |D_S W_0|_{\zeta=0}|^2 (\epsilon_{\rho}-\frac{\epsilon_s}{4}+2\,\epsilon_{\tau_s})
+\frac{(\alpha q_{\rho})^2}{4\pi}\frac{\rho^2}{(s+h\rho) }\, \epsilon_{\tau_s}\right. \nn \\
&& \left.
-2 s (b+1/(2s)) B \left[ \cos[b(C_0+h \psi_{\rho})] {\rm Re}(D_S W_0|_{\zeta=0})
-\sin [b(C_0+h \psi_{\rho})] {\rm Im}(D_S W_0|_{\zeta=0})\right]e^{-b (s+h\rho)}\right. \nn \\
&& \left. + \frac 32 \left[{\rm Re}(W_0) {\rm Re}(D_S W_0|_{\zeta=0})+{\rm Im}(W_0) {\rm Im}(D_S W_0|_{\zeta=0})\right] \epsilon_s\right. \nn \\
   && \left. + \frac{1}{2s}\left[\frac 83  (a A)^2  \sqrt{\tau_s}\, \tau_b^{3/2}\,e^{-2 a \tau_s}
   + 4 a A W_0 \cos\left(a \psi_s\right) \tau_s\, e^{-a \tau_s} +W_0^2 \frac{3 \epsilon_s- \epsilon_{\rho}}{4}  \right.\right. \nn \\
   && \left.\left. + \frac{2}{\alpha}(h b B)^2 \tau_b^{3/2}  e^{-2 b (s+ h \rho)}
   +2 h b B W_0 \cos\left[ b \left(C_0 +h \psi_{\rho}\right)\right] \rho \,e^{-b (s +h \rho)}\right] \right\}.
\label{VO3}
\eea
The potential at order $\vo^{-2}$ depends on four fields: $\tau_b$, $s$, $C_0$ and $\rho$.
However, the minimisation with respect to $S$ and $\rho$ implies that:
\be
\langle D_S W|_{\zeta=0}\rangle=0\qquad\text{and}\qquad\langle\rho\rangle=0\,,
\ee
leaving a flat potential for $\tau_b$ and justifying our expansion of the
K\"ahler potential around the singularity obtained by shrinking the blow-up mode $\rho$.
The potential at order $\vo^{-3}$ then reads:
\be
V_{\mc{O}(\vo^{-3})}=\frac{1}{2 \langle s\rangle}
\left[\frac 83  (a A)^2  \sqrt{\tau_s}\, \frac{e^{-2 a \tau_s}}{\tau_b^{3/2}}
   - 4 a A W_0 \tau_s\frac{e^{-a \tau_s}}{\tau_b^3} + \frac{3}{4} \frac{\zeta \langle s\rangle ^{3/2}}{\tau_b^{9/2}} W_0^2
   + \frac{2}{\alpha}(h b B)^2 \frac{e^{-2 b \langle s\rangle}}{\tau_b^{3/2}}\right] .
\label{VO3s}
\ee
where we have already minimised with respect of $\psi_s$ and we have set $s=\langle s\rangle = 1/g_s$.
Notice that the leading order stabilisation of the blow-up mode $\rho$ at $\langle\rho\rangle=0$
eliminates the $\rho$ dependence in the exponentials and the last term in (\ref{VO3}).
This is important because this is the only extra term that could give a negative contribution to the scalar potential.
Considering the scalar potential for $\rho$ by adding the $\rho$-dependent terms in (\ref{VO3})
to the D-term, the VEV of $\rho$ is slightly moved away from the singularity
but the minimum is at a value $\langle\rho\rangle\sim 1/\vo$
inducing a much suppressed contribution to the scalar potential of order
$\delta V_F\sim 1/\vo^4$, and therefore can be safely neglected.

Similarly for $(D_S W_0|_{\zeta=0})$, this quantity has been fixed
at $\langle D_S W|_{\zeta=0}\rangle=0$ only focusing on the potential at order $\vo^{-2}$.
The leading order correction to this result comes from considering also the
dilaton dependent terms in (\ref{VO3}). They slightly move the minimum to
$\langle D_S W|_{\zeta=0}\rangle\sim 1/\vo$ giving rise again to contributions of the order
$\delta V_F\sim 1/\vo^4$, which can therefore be safely neglected.

What we are left with then, is a potential of the standard LVS form (\ref{Vlvs})
plus an additional positive definite term coming from the non-perturbative effects at the singularity:
\be
V = V_{\rm LVS} + V_{\rm up}\,,
\ee
where:
\be
V_{\rm up} \propto h^2\,\frac{e^{-2b\langle s\rangle}}{\vo}\,,
\label{uplift2}
\ee
with positive proportionality factor. This is precisely of the form (\ref{uplift1}), with $\alpha=5/3$ and the warp factor substituted by a very similar expression now as an exponential of the dilaton field which is fixed by fluxes. Therefore the effect of this term is identical to a large warping for weak coupling strings. Notice also that the uplifting term (\ref{uplift2}) is proportional to $h$,
and so it goes to zero for $h=0$.

Let us now minimise the scalar potential with respect to $\tau_s$ by solving $\partial V/\partial \tau_s=0$.
In the limit $a\tau_s\gg 1$, we find:
\be
e^{-a \tau_s}= \frac{3\sqrt{\tau_s}}{4 a A}\frac{W_0}{\tau_b^{3/2}}\sim\mc{O}(\vo^{-1})
\quad\Rightarrow\quad a \tau_s =\ln\left(\frac{4 a A}{3\sqrt{\tau_s}}\right)+\ln\left(\frac{\tau_b^{3/2}}{W_0}\right)
\simeq\ln\left(\frac{\tau_b^{3/2}}{W_0}\right)\,.
\label{tsVEV}
\ee
Substituting this result back in (\ref{VO3s}) we end up with the following effective potential:
\be
V_{\mc{O}(\vo^{-3})}=\frac{3}{4\langle s\rangle}  \frac{W_0^2}{\tau_b^{9/2}} \left\{\frac{\zeta \langle s\rangle ^{3/2}}{2}
-\left[\frac{\ln\left(\tau_b^{3/2}/W_0\right)}{a}\right]^{3/2}
   + \frac{4}{3\alpha}\left(\frac{h b B}{W_0}\right)^2 e^{-2 b \langle s\rangle}\tau_b^3 \right\}.
\label{VO3tb}
\ee
The solution of $\partial V/\partial \tau_b=0$ is then:
\be
\frac{\zeta \langle s\rangle ^{3/2}}{2}=\left[\frac{\ln\left(\tau_b^{3/2}/W_0\right)}{a}\right]^{3/2}
\left(1-\frac{1}{2\ln\left(\tau_b^{3/2}/W_0\right)}\right)-\frac{4}{9\alpha}
\left(\frac{h b B}{W_0}\right)^2 e^{-2 b \langle s\rangle}\tau_b^3\,.
\label{tbVEV}
\ee
This implicitly fixes the volume $\vo$ or $\tau_b$ as a function of $\langle s \rangle, W_0, B$ and the constants $h,a, b,\alpha$. Substituting this result back in (\ref{VO3tb}) we find that the value of the potential at the
minimum is: \footnote{Explicit string calculations at one-loop order have led to the need to redefine the K\"ahler modulus that corresponds to the proper chiral superfield in the supergravity action \cite{Redef}. This field redefinition is model dependent but has been found to be needed for blow-up modes for orientifolds and when D3 and D7-branes are present at a singularity. It is then important to re-analyse our results of this section taking into account  that the blow-up field is subject to a field redefinition. This analysis is carried out in appendix \ref{AppFieldRedef} where we show that a field redefinition does not qualitatively change our results.}
\be
\langle V_{\mc{O}(\vo^{-3})}\rangle=\frac{3}{4\langle s\rangle}\frac{W_0^2}{\tau_b^{9/2}}
\left\{-\frac{\left[\ln\left(\tau_b^{3/2}/W_0\right)\right]^{1/2}}{2 a^{3/2}}
+\frac{8}{9\alpha}
\left(\frac{h b B}{W_0}\right)^2 e^{-2 b \langle s\rangle}\tau_b^3\right\}.
\label{VO3min}
\ee
Notice that in the absence of the non-perturbative quiver superpotential ($B=0$) this is negative giving rise to the standard AdS vacuum.
For $B\neq 0$ the second term, being positive, lifts the minimum allowing the possibility of having a de Sitter minimum and even a destabilisation of the minimum for large enough values.

\subsubsection{`Tuning' of the cosmological constant}

Given that the parameters $\langle s\rangle, W_0, B$ are determined by the fluxes and
the VEVs of hidden sector matter fields,
they can be adjusted to cancel the vacuum energy up to this order.
In order to find a Minkowski vacuum we need therefore to perform the following tuning:
\be
\frac{\left[\ln\left(\tau_b^{3/2}/W_0\right)\right]^{1/2}}{2 a^{3/2}}=
\frac{8}{9\alpha}
\left(\frac{h b B}{W_0}\right)^2 e^{-2 b \langle s\rangle}\tau_b^3\,.
\label{tuning}
\ee

Let us now try to estimate the amount of tuning needed to get a Minkowski vacuum,
by considering all the underlying parameters fixed except for $B$.
Hence we have
three equations, (\ref{tsVEV}), (\ref{tbVEV}) and (\ref{tuning}), in three unknowns,
$\tau_s$, $\tau_b\simeq \vo^{2/3}$ and $B$ whose solution is:
\be
\langle\tau_s\rangle\simeq \langle s \rangle \left(\frac{\zeta}{2}\right)^{2/3},
\quad\langle\vo\rangle \simeq W_0 \,e^{a\langle\tau_s\rangle},
\quad B\simeq\underset{\mc{O}(1)}{\underbrace{\left(\frac{3\langle\tau_s\rangle^{1/4}}{4 h b}\sqrt{\frac{\alpha}{a}}\right)}}
 \, \left(\frac{\langle\vo\rangle}{W_0}\right)^{\frac{b}{a} \left(\frac{2}{\zeta}\right)^{2/3}-1}\,.
\label{sol}
\ee
Hence the order of magnitude of $B$ depends crucially on $\zeta$ (and so the Euler number of the underlying Calabi-Yau) and the choice of parameters $a$ and $b$.

Given a Minkowski vacuum, one can obtain a discretuum of de Sitter vacua with small cosmological constant by varying $W_0$ which depends on all flux quanta.
This can be seen by considering a small deviation $\epsilon$ in the relation (\ref{tuning}),
which would result in a vacuum energy of the form:
\be
\langle V_{\mc{O}(\vo^{-3})}\rangle=\frac{3}{4\langle s\rangle}\frac{W_0^2}{\tau_b^{9/2}}\,\epsilon\,.
\ee
We have then achieved in the LVS the same effect as anti D3-branes. In fact,
we showed that D3/E(-1) non-perturbative effects at singularities provide an uplifting term that
allows an almost continuum tuning of the vacuum energy by small changes in the fluxes. This is in contrast with proposals for D-term uplift in which the needed substantial warping was not explicitly achieved and may require topological conditions on the internal manifold difficult to satisfy \cite{Dterm}. The main advantage over anti D3-brane constructions is that in this case the effective field theory is under control especially since it is manifestly supersymmetric.

\subsubsection{Generation of non-perturbative superpotentials at singularities}
\label{generation}

We now discuss how to obtain a non-zero prefactor
of the dilaton-dependent non-perturbative superpotential in a way compatible with the stabilisation
of the blow-up mode $\rho$ at the singular locus by imposing the vanishing of the Fayet-Iliopoulos term.
We can envisage three different scenarios:
\begin{itemize}
\item In the presence of just one anomalous $U(1)$, one would require
matter fields with opposite charges to have VEVs such that their contribution to the D-term vanishes
consistent with a non-zero value of the prefactor $B$.

\item In the more general case with more than one anomalous $U(1)$ and multiple blow-up modes,
a non-vanishing prefactor can be obtained without necessarily requiring cancellation of
matter field contributions to all the D-terms.
An explicit example of this was studied by the Romans \cite{Romans}. The construction involved
a $\mathbb{Z}_5$ quiver with gauge group $SU(5)\times U(1)_5 \times U(1)_1$
(and so two anomalous $U(1)$'s) and two blow-up modes $Q_1$ and $Q_2$.
The non-perturbative superpotential at the singularity is given by:
\be
W_{\rm np} = B \, e^{-b (S+a_1 Q_1+a_2 Q_2)}\,,
\ee
with the prefactor $B$ depending on open string fields $B= \phi/Z^2$,
where $\phi$ is charged only under $U(1)_1$ and $Z$ is a composite field
built out of fields charged under both anomalous $U(1)$'s.
Given that the combination of blow-ups entering in $W_{\rm np}$ is the same
as that entering in the FI-term of $U(1)_5$, $\xi_5 \propto a_1 \rho_1+a_2 \rho_2$,
the D-term associated with this $U(1)$
can fix exactly this combination to zero (in accord with our result in the previous section)
without having to require cancellations between matter fields VEVs.
On the other hand, the D-term of $U(1)_1$ fixes
\be
|\phi|^2 = \xi_1 \propto b_1 \rho_1+b_2 \rho_2 = m^2 \neq 0\,.
\ee
Notice that this stabilisation would leave a flat direction which could then be
fixed at non-zero value by the inclusion of a tachyonic
susy-breaking mass term for $\phi$ (justifying our assumption of $m \neq 0$).
Hence the prefactor $B$ is non-zero since it turns out to be $B=m/Z_{\rm light}^2$
where in the denominator $Z_{\rm light}$ depends only on the fields which remain light.
The composite field $Z_{\rm light}$ then corresponds to the standard run-away
direction of ADS-like superpotentials in the global supersymmetry case where the potential is proportional to $|\partial W/\partial Z|^2 $  \cite{seiberg}.
This direction could be fixed at non-zero value by
supergravity effects if there is a finite solution to $D_ZW=0$. For a discussion along these lines see for instance
\cite{Florea}. Furthermore soft breaking terms will also provide contributions to the the $Z$ potential and generically lifting the runaway behaviour.

\item The prefactor $B$ might not depend at all on matter fields for orientifold projections
such that $h^{1,1}_-\neq 0$ as discussed in \cite{GKPW}.
\end{itemize}

\subsection{De Sitter LVS for positive Euler numbers}

It is well known that the standard realisation of the LARGE Volume Scenario
with non-perturbative effects in the geometric regime and $\alpha'$ corrections
relies on the fact that the Calabi-Yau manifold has negative Euler number $\chi$
(or positive coefficient $\zeta \propto -\chi$).
This means that the number of complex structure moduli ($h^{1,2}$)
is larger than the number of K\"ahler moduli ($h^{1,1}$).
This amounts to essentially half of all Calabi-Yau manifolds because of mirror symmetry.

However ref. \cite{CMV} opens up the possibility to obtain LVS also for $\chi\geq 0$
(or equivalently $\zeta\leq 0$). In fact, the authors of \cite{CMV} pointed
out that, in order to avoid the shrinking of the four-cycle supporting chiral matter,
one has to wrap the visible sector branes on intersecting rigid divisors.
In this way, in the absence of singlets which can get non-zero VEVs without
breaking any of the visible sector gauge groups, one can
perform an appropriate choice of brane set-up and world-volume fluxes which
leads to the D-term stabilisation of all these rigid divisors except for
one. This remaining flat direction, which we shall denote $\tau_{\rm vs}$
since its size is constrained to be small by the requirement of obtaining a
visible sector gauge coupling of the correct size, can be fixed by the
inclusion of string loop corrections \footnote{For a similar mechanism
to fix the visible sector four-cycle via string loops but in the
presence of non-vanishing singlet VEVs see \cite{ccq2}.} to be proportional to $\tau_s$:
$\langle\tau_{\rm vs}\rangle\sim\langle\tau_s\rangle$.
The final contribution to the scalar potential looks like \cite{CMV}:
\be
V_{\rm loop}^{({\rm s})}\simeq\frac{c_{\rm loop}^{({\rm s})} W_0^2}{\vo^3 \sqrt{\tau_s}}\,,
\label{loop}
\ee
where $c_{\rm loop}^{({\rm s})}$ is an unknown coefficient which depends on the complex structure moduli fixed at tree level.
Given that $\tau_s \sim \mc{O}(10)$ this term scales as $1/\vo^3$, and so for $c_{\rm loop}^{({\rm s})}>0$,
it has the potentiality to give rise to an AdS minimum at exponentially large volume even if $\zeta\leq 0$.
We need however to check that the tuning of $c_{\rm loop}^{({\rm s})}$ needed to obtain such a minimum
does not lead us to a regime where we cannot trust the perturbative expansion anymore.
As explained in \cite{ccq1}, the parameter that controls this expansion is:
\be
\epsilon_{\rm loop}^{({\rm s})}\equiv\frac{c_{\rm loop}^{({\rm s})}}{\tau_s}\ll 1\,,
\label{expansionparams}
\ee
which comes from the expansion of the inverse one-loop corrected K\"ahler metric.
Notice that we expect two-loop contributions to the scalar potential
to be suppressed by additional loop factors of $(16\pi^2)$, and so
these higher-loop corrections can be neglected if $\epsilon_{\rm loop}^{({\rm s})}\ll 1$.

On top of these loop corrections, we have also $g_s$ effects coming from loops
of open string states living on the large cycle $\tau_b$:
\be
V_{\rm loop}^{({\rm b})}\simeq\frac{c_{\rm loop}^{({\rm b})} W_0^2}{\vo^{10/3}}\,,
\label{Looplarge}
\ee
and in this case the parameter that controls the perturbative expansion is:
\be
\epsilon_{\rm loop}^{({\rm b})}\equiv\frac{c_{\rm loop}^{({\rm b})}}{\tau_b}\simeq \frac{c_{\rm loop}^{({\rm b})}}{\vo^{2/3}}\ll 1\,.
\label{expansionparamb}
\ee

Let us start by neglecting the string loop correction (\ref{Looplarge}) due to their volume suppression.
The scalar potential takes the form (neglecting the prefactor):
\be
V=\mu_1  \sqrt{\tau_s}\, \frac{e^{-2 a \tau_s}}{\vo}
   - \mu_2 W_0 \tau_s\frac{e^{-a \tau_s}}{\vo^2} - \mu_3 \frac{W_0^2}{\vo^3}
   + \frac{c_{\rm loop}^{({\rm s})} W_0^2}{\vo^3 \sqrt{\tau_s}} .
\label{VO3geom}
\ee
where:
\be
\mu_1\equiv\frac 83  (a A)^2>0\,,\qquad \mu_2 \equiv 4 a A>0 \qquad\text{and}\qquad
\mu_3\equiv \frac 34 \frac{|\zeta|}{g_s^{3/2}}\geq 0\,.
\ee
Solving $\partial V/\partial \tau_s=0$ in the limit $a\tau_s\gg 1$ gives:
\be
e^{-a\tau_s}\simeq\frac{3\sqrt{\tau_s}}{8 a A}\frac{W_0}{\vo}\left[1\pm \left(1-\frac{ c_{\rm loop}^{({\rm s})}}{3 a\tau_s^3}\right)\right]
\quad\Rightarrow\quad \tau_s \simeq\frac{\ln \vo}{a}\quad \text{for}\quad W_0\sim \mc{O}(1)\,.
\label{compare}
\ee
A careful analysis considering also the second derivative with respect to $\tau_s$
shows that in order to get a minimum we need to take the solution with the positive sign.
Notice that in the case with $c_{\rm loop}^{({\rm s})}=0$ this solution reduces to (\ref{tsVEV}).
Substituting this result back in (\ref{VO3geom}) we find (again in the limit $a\tau_s\gg 1$):
\be
V\simeq \frac 32 \frac{W_0^2}{\vo^3}\left[
   - \frac{\left(\ln \vo\right)^{3/2}}{a^{3/2}}
   - \frac{|\zeta|}{2 g_s^{3/2}}
   + \frac{2 \sqrt{a} \,c_{\rm loop}^{({\rm s})}}{3\sqrt{\ln \vo}} \right] .
\label{VO3geoms}
\ee
The solution of $\partial V/\partial \vo=0$ is then:
\be
\frac{\ln\vo}{a}= \sqrt{\frac{2\,c_{\rm loop}^{({\rm s})}}{3}} \left(1-\epsilon\right)^{1/2}\simeq \sqrt{\frac{2\,c_{\rm loop}^{({\rm s})}}{3}}\,,
\label{tbVEVgeom}
\ee
where:
\be
\epsilon \equiv \frac{3 |\zeta| g_s^{-3/2}}{4 c_{\rm loop}^{({\rm s})}}\sqrt{\frac{\ln\vo}{a}}
\simeq \frac{|\zeta|}{2} \left(\frac 32\right)^{3/4}\left(\frac{g_s^{-2}}{c_{\rm loop}^{({\rm s})}}\right)^{3/4}\ll 1
\quad \text{for}\quad c_{\rm loop}^{({\rm s})}\gg g_s^{-2}\,.
\ee
Taking this value of $c_{\rm loop}^{({\rm s})}$ and comparing (\ref{tbVEVgeom}) with (\ref{compare}),
we realise that the minimum is at:
\be
\langle\tau_s\rangle\simeq \sqrt{\frac{2\,c_{\rm loop}^{({\rm s})}}{3}}\sim g_s^{-1}\sim\mc{O}(10)
\qquad\text{and}\qquad\langle\vo\rangle\sim W_0 \,e^{a \langle\tau_s\rangle}\gg 1\,.
\ee
However this minimum is not in a regime where we can trust the perturbative expansion
since the parameter (\ref{expansionparams}) becomes:
\be
\epsilon_{\rm loop}^{({\rm s})}\sim \sqrt{c_{\rm loop}^{({\rm s})}}\sim g_s^{-1}\gg 1\qquad\text{for}\qquad
g_s\ll 1\,.
\ee
This implies that the string loop correction (\ref{loop}) can never beat the $\alpha'$
correction and yield a minimum in a regime where we can trust the effective field theory.
However this is not the case for the loop correction (\ref{Looplarge}) coming from the large cycle.
In fact, setting $c_{\rm loop}^{({\rm s})}=0$ without loss of generality,
the effective potential in terms of $\vo$ reads:
\be
V\simeq \frac 32 \frac{W_0^2}{\vo^3}\left[
   - \frac{\left(\ln \vo\right)^{3/2}}{a^{3/2}}
   - \frac{|\zeta|}{2 g_s^{3/2}}
   + \frac{2\,c_{\rm loop}^{({\rm b})}}{3\vo^{1/3}} \right] .
\label{VO3geomsnew}
\ee
The solution of $\partial V/\partial \vo=0$ is then:
\be
\frac{\ln\vo}{a}= \left(\frac{20}{27}\right)^{2/3}\left(\frac{c_{\rm loop}^{({\rm b})}}{\vo^{1/3}}\right)^{2/3}
\left(1-\delta\right)^{2/3}\simeq \left(\frac{20}{27}\right)^{2/3}\left(\frac{c_{\rm loop}^{({\rm b})}}{\vo^{1/3}}\right)^{2/3}\,,
\label{tbVEVgeomnew}
\ee
where:
\be
\delta \equiv \frac{27 |\zeta| }{40 }\frac{g_s^{-3/2} \vo^{1/3}}{c_{\rm loop}^{({\rm b})}} \ll 1
\qquad \text{for}\qquad c_{\rm loop}^{({\rm b})}> g_s^{-3/2}\vo^{1/3}\,.
\ee
Given that all the known Calabi-Yau three-folds have $1/2\lesssim|\zeta|\lesssim 3/2$,
the condition $\delta\ll 1$ is well satisfied by taking $c_{\rm loop}^{({\rm b})}\simeq 4\, g_s^{-3/2}\vo^{1/3}$
since it gives $0.08\lesssim\delta\lesssim 0.25$.
In this case the parameter (\ref{expansionparamb}) that controls the perturbative expansion
is still smaller than unity since it turns out to be volume suppressed:
\be
\epsilon_{\rm loop}^{({\rm b})}\sim \frac{4\, g_s^{-3/2}}{\vo^{1/3}}\ll 1\qquad \text{for}\qquad
\vo\gg 1\,.
\ee
This minimum is AdS but it can be turned into a dS vacuum due to the
positive contribution (\ref{uplift2}) coming from D3/E(-1) non-perturbative effects at singularities
in the same way described in the previous sections.

\section{Phenomenological implications}
\label{Implications}

Even though the mechanism we have used for getting de Sitter vacua is very different from
anti D3-branes at a warped throat, once uplifting has been achieved,
the main physical implications are very similar to the standard LVS.
However, they can provide a more robust origin to the different implications of LVS.

\begin{itemize}
\item{} {\it Inflation:} Models of K\"ahler moduli inflation \cite{KahlerInflation} are based on the LVS and
depend very much on the positive uplift term to get early Universe accelerated slow roll. A criticism to these scenarios could be that the most important
contribution to the almost de Sitter expansion relies in the uplifting mechanism that generically requires the introduction of anti D3-branes.
However, if this term comes from a manifestly supersymmetric theory, as we have here, instead of a non linear realisation of supersymmetry, it
makes the status of these models more robust. This is what can be obtained from our mechanism since the whole dynamics is identical to the existing models of inflation except for the origin of the
`uplift' term. \footnote{We want to emphasise that uplift term is not a proper terminology in our mechanism since the term appears on equal footing to the other terms in the scalar potential and we have explicitly found the minimum of the full potential.}

\item{} {\it Supersymmetry breaking:}
The new mechanism for obtaining dS vacua can play an important r\^ole in the study of soft supersymmetry breaking. There are several realisations of the soft breaking terms within the LVS. This depends on whether the cycle supporting the Standard Model brane participates or not in the breaking of supersymmetry. If it does, the soft terms are of order the gravitino mass $m_{3/2}\propto 1/\vo$ \cite{cqs}. In the general expression for the soft scalar masses there appears a contribution from the vacuum energy $m_0^2= V_0 +\cdots$. Since the uplifting term is of order $1/\vo^3$ its contribution to scalar masses is much smaller than the gravitino mass and therefore does not play any r\^ole on the soft breaking Lagrangian.
However if the Standard Model cycle does not contribute to supersymmetry breaking, then the soft terms can be hierarchically smaller than the gravitino mass with scalar masses of order $m_0\sim 1/\vo^{3/2}, 1/\vo^2$ \cite{bckmq} (see however \cite{joepedro} when field redefinitions are needed). In this case the uplifting term is crucial for determining the soft scalar masses \cite{shanta1}. Given that our scenario provides an explicit supersymmetric source for obtaining dS vacua, the contribution to the scalar masses can be computed in a reliable way. In particular what we have is an `F-term uplift'
realised via the F-term of a blow-up mode whose contribution to the soft masses in the Standard Model brane is negligible since the visible sector is
localised at a different singularity. Hence the leading order contribution to the soft masses comes from the F-term of the dilaton which is
of order $1/\vo^2$, allowing for a realisation of the large gravitino mass scenario of \cite{bckmq,shanta2} (see also \cite{joe}).

\item{} {\it An AdS/CFT dual description?:}
Our uplift potential takes the form:
\be
V_{\rm up}\simeq \frac{e^{-2 b s}}{ \vo^{\alpha}},
\label{Vup}
\ee
with $\alpha=1$ for the case with $h\neq0$ and $\alpha=2$ for $h=0$ and $s$ fixed by ratios of integers from fluxes. This has a very similar structure as the anti D3-brane potential with  $e^{-2bs}$ playing the r\^ole of the minimal value of the warp factor.
The term (\ref{Vup}) can arise from a non-perturbatively generated superpotential of a gauge theory; one thus might ask whether there is a relation to a dual  theory along the lines of \cite{k1}.
On the other hand there seem to be some key differences between our system and a warped throat with an anti-brane. Firstly, supersymmetry is broken by a bulk modulus; unlike the anti-brane case where the brane is the source of supersymmetry breaking. Furthermore, the value at which $s$ is fixed is sensitive to the flux quanta in all the three-cycles in the Calabi-Yau; this makes it difficult to associate $e^{-2bs}$ with the infrared scale of a throat. It may be interesting to have a proper understanding of this in terms of AdS/CFT duality.
\end{itemize}

\section{Conclusions}
\label{Conclusions}

Obtaining de Sitter space solutions from string theory is hard but very relevant for phenomenology. Despite several attempts, so far the most convincing proposals have involved anti D3-branes which are required to provide a positive source of potential energy that combines with a concrete mechanism of modulus stabilisation to lift an AdS minimum
to positive values. Moreover the amount of this lifting is controlled in an almost continuous manner given the fact that this contribution comes with an exponential  warped factor depending on ratios of integers determined by fluxes.

The fact that this mechanism is not explicitly supersymmetric has raised criticism and doubts on how reliable and stringy this mechanism is. Even though these doubts may not be fully justified, it is reassuring to identify alternative mechanisms that achieve the same results but coming from a fully supersymmetric effective action. We regard this to be the main result of this article.

We achieved de Sitter vacua from a combination of non-perturbative effects. The first is the standard one coming from either gaugino condensation on D7-branes or Euclidean D3 instantons wrapping rigid four-cycles in the geometric regime. The new element is to include dilaton-dependent non-perturbative effects arising from either gaugino condensation on space-time filling D3-branes or E(-1)-instantons at singularities. A combination of both is what gives rise to a tunable (as in the anti D3-branes) positive vacuum energy. This makes possible fully supersymmetric treatments of inflation and soft supersymmetry breaking in the LVS making the calculations more controllable and the scenarios more robust.
Moreover these dilaton-dependent non-perturbative superpotentials, when combined with string loop effects,
open up the possibility of realising new LVS for manifolds with zero or positive Euler number.

Open questions are legion: a detailed calculation of soft terms for different scenarios of supersymmetry breaking, explicit realisations of this scenario in compact Calabi-Yau models following the recent constructions in \cite{CMV},  explicit calculations of the next order corrections to the scalar potential (of order $1/\vo^4$), etc. We hope to report on some of these issues in the near future.

\section*{Acknowledgments}

We thank Matteo Bertolini, Massimo Bianchi, Shanta de Alwis, Thomas Grimm, Sven Krippendorf,
Liam McAllister, Jose F. Morales, Marco Serone, Gary Shiu and in particular Joe Conlon and Roberto Valandro for useful discussions.
We acknowledge DAMTP-CTC, Cambridge and ICTP, Trieste where part of this work was done.
We thank the organisers of PASCOS 2011, Cambridge for
providing a good atmosphere to develop this project. AM is funded by the EU and the University of Cambridge.
CB's research was supported in part by funds from the Natural Sciences and Engineering Research Council (NSERC) of Canada. Research at the Perimeter Institute is supported in part by the Government of Canada through Industry Canada, and by the Province of Ontario through the Ministry of Research and Information (MRI).
FQ also wishes to thank the hospitality of KITPC during the final stages of this project.
This research was partly supported by the Project of Knowledge Innovation
Program (PKIP) of Chinese Academy of Sciences, Grant No. KJCX2.YW.W10.

\appendix

\section{Effect of field redefinition on the potential}
\label{AppFieldRedef}

In this appendix we shall study the effect of a field redefinition of the blow-up mode $\rho$
due to one-loop corrections to the gauge kinetic function \cite{Redef}.
This redefinition takes the form $\rho\to\rho-\beta\ln\vo$ where the coefficient $\beta$ is:
\be
\beta=\frac{b_0}{12\pi h}\,,
\label{beta}
\ee
where $b_0$ is the one-loop $\beta$-function coefficient
The volume of the blow-up mode is not given by $\rho$ anymore but by $(\rho-\beta\ln\vo)$,
and so the quiver locus corresponds to $\rho=\beta\ln\vo$.

We shall work at fixed dilaton.
The K\"ahler potential, in the large volume limit $\tau_b\gg\tau_s$, can be expanded as
(keeping only terms up to $\mc{O}(\vo^{-1})$):
\be
K\simeq -3\ln \tau_b +2\epsilon_{\tau_s}- \frac{\zeta\langle s\rangle^{3/2}}{\tau_b^{3/2}}
+\alpha \frac{(\rho-\frac{3\beta}{2}\ln\tau_b)^2}{\tau_b^{3/2}}\,.
\label{KPotRedef}
\ee
On the other hand the superpotential takes the form:
\be
W=W_0+ A\, e^{-a T_s}+B e^{-b \left(\langle S\rangle+h Q\right)}\,.
\ee
The K\"ahler metric is a symmetric matrix which reads (with the leading order correction in a large volume expansion):
\be
K_{I\bar{J}}=\left(
\begin{array}{ccc}
 \frac{3}{4
   \tau_b^2}\left(1+\frac{5\epsilon'_{\rho}-5\epsilon_s}{4} \right) & -\frac{9  \sqrt{\tau_s}}{8 \tau_b^{5/2}}
   &  -\frac{3 \alpha }{4 \tau_b^{5/2}}\left(\rho-\frac{3\beta}{2}\ln\tau_b+\beta\right) \\
K_{1\bar{2}} & \frac{3 }{8\tau_b^{3/2} \sqrt{\tau_s}} & 0 \\
K_{1\bar{3}} & 0 & \frac{\alpha }{2 \tau_b^{3/2}}
\end{array}
\right). \nn
\ee
The inverse K\"ahler metric with the leading order correction in a large volume expansion reads:
\be
K^{I\bar{J}}=\left(
\begin{array}{ccc}
 \frac{4}{3} \tau_b^2 \left(1+\frac{18\epsilon_{\tau_s}+5\epsilon_s+\epsilon'_{\rho}}{4} \right) &
 4 \tau_b \tau_s \left(1+\frac{18\epsilon_{\tau_s}+5\epsilon_s+\epsilon'_{\rho}}{4} \right) &
 2 \tau_b \left(\rho-\frac{3\beta}{2}\ln\tau_b+\beta \right)\left(1+\frac{18\epsilon_{\tau_s}+5\epsilon_s+\epsilon'_{\rho}}{4} \right) \\
 K^{1\bar{2}} & \frac 83 \tau_b^{3/2} \sqrt{\tau_s} \left(1+\frac 92 \epsilon_{\tau_s} \right) &
6 \tau_s \left(\rho-\frac{3\beta}{2}\ln\tau_b+\beta \right)\left(1+\frac{18\epsilon_{\tau_s}+5\epsilon_s+\epsilon'_{\rho}}{4} \right) \\
 K^{1\bar{3}} & K^{2\bar{3}} & \frac{2\tau_b^{3/2}}{\alpha}\left(1+\frac 32 \epsilon'_{\rho}\right)
\end{array}
\right). \nn
\ee
where:
\be
\epsilon'_{\rho}\equiv \frac{\alpha \left( \rho-\frac{3\beta}{2}\ln\tau_b\right)^2}{\tau_b^{3/2}}
\sim \mc{O}(\vo^{-1})\ll 1\,.
\ee
By turning on a flux on a two-cycle inside the four-cycle whose volume is given by the blow-up mode $(\rho-\beta\ln\vo)$,
the D-term potential at order $\vo^{-2}$ forces the shrinking of this divisor at the singularity
since it fixes $\rho$ at:
\be
K_Q = \frac{\alpha  }{\tau_b^{3/2}}\left(\rho-\frac{3\beta}{2}  \ln \tau_b \right)=0 \quad \Leftrightarrow\quad
\langle\rho\rangle=\frac{3\beta}{2}  \ln \tau_b\,.
\label{DminRedef}
\ee
The leading order scalar potential at order $\vo^{-3}$ is then
(after minimising with respect to the axions $\psi_s$ and $\psi_{\rho}$):
\bea
V&=&\frac{1}{2 \langle s\rangle \tau_b^3} \left\{
\frac 83 (a A)^2 \tau_b^{3/2} \sqrt{\tau_s} e^{-2 a \tau_s}- 4 a A W_0 \tau_s\, e^{-a \tau_s}
   + \frac{2}{\alpha}(h b B)^2 \tau_b^{3/2}  e^{-2 b (\langle s\rangle+ h \langle\rho\rangle)}\right. \nn \\
   && \left. -2 h b B W_0 \left(\langle\rho\rangle-\frac{3\beta}{2}\ln\tau_b+3\beta\right)
   \,e^{-b (\langle s\rangle +h \langle\rho\rangle)} \right. \nn \\
   && \left. +\frac{W_0^2}{4\tau_b^{3/2}}\left[3 \zeta \langle s\rangle^{3/2} - \alpha  \left(\langle\rho\rangle-\frac{3\beta}{2}\ln\tau_b\right) \left(\langle\rho\rangle-\frac{3\beta}{2}\ln\tau_b+24 \beta \right)\right]\right\}\,. \nn
\eea
Substituting the value for $\langle\rho\rangle$ found in (\ref{DminRedef}),
the scalar potential simplifies to:
\bea
V_{\mc{O}(\vo^{-3})}&=&\frac{1}{2 \langle s\rangle}
\left[\frac 83  (a A)^2  \sqrt{\tau_s}\, \frac{e^{-2 a \tau_s}}{\tau_b^{3/2}}
   - 4 a A W_0 \tau_s\frac{e^{-a \tau_s}}{\tau_b^3} + \frac{3}{4} \frac{\zeta \langle s\rangle ^{3/2}}{\tau_b^{9/2}} W_0^2 \right. \nn \\
   && \left. + \frac{2}{\alpha}(h b B)^2 \frac{e^{-2 b \langle s\rangle}}{\tau_b^{\frac 32(1+2 h b \beta)}}
   \left(1-\frac{3\alpha W_0}{h b B} \,\beta\,e^{b \langle s\rangle} \tau_b^{-\frac 32 (1- h b \beta)}\right)\right] .
\label{VO3sRedef}
\eea
Notice that this expression differs from (\ref{VO3s}) only for the last term proportional to $\beta$.
Given that this extra term has an overall negative sign coming from the minimisation with respect to the axion $\psi_{\rho}$,
it could be dangerous to find a dS vacuum if it dominates over the term proportional to $e^{-2 b \langle s \rangle}$.
We shall however now show that this is never the case.

Given that the new term does not depend on $\tau_s$ we find the same condition (\ref{tsVEV}) from the
minimisation with respect to $\tau_s$ which substituted in (\ref{VO3sRedef}) gives:
\bea
V_{\mc{O}(\vo^{-3})}&=&\frac{3}{4\langle s\rangle}  \frac{W_0^2}{\tau_b^{9/2}} \left\{\frac{\zeta \langle s\rangle ^{3/2}}{2}
-\left[\frac{\ln\left(\tau_b^{3/2}/W_0\right)}{a}\right]^{3/2} \right. \nn \\
 && \left.  + \frac{4}{3\alpha}\left(\frac{h b B}{W_0}\right)^2 e^{-2 b \langle s\rangle}\tau_b^{3(1- h b \beta)}
   \left(1-\frac{3\alpha W_0}{h b B} \,\beta\,e^{b \langle s\rangle} \tau_b^{-\frac 32 (1- h b \beta)}\right)\right\}.
\label{VO3tbRedef}
\eea
The solution of $\partial V/\partial \tau_b=0$ is then:
\bea
\label{taubVEVRedef}
\frac{\zeta \langle s\rangle ^{3/2}}{2}&=&\left[\frac{\ln\left(\tau_b^{3/2}/W_0\right)}{a}\right]^{3/2}
\left(1-\frac{1}{2\ln\left(\tau_b^{3/2}/W_0\right)}\right) \\
&& -\frac{4}{9\alpha}
\left(\frac{h b B}{W_0}\right)^2 \left(1+ 2 h b \beta\right)\,e^{-2 b \langle s\rangle}\tau_b^{3 (1- h b \beta)}
\left(1-\frac{3\alpha W_0}{h b B}
\frac{2+ h b \beta}{1+ 2 h b \beta}\,\beta\,e^{b \langle s\rangle} \tau_b^{-\frac 32 (1- h b \beta)}\right)\,. \nn
\eea
Substituting this result back in (\ref{VO3tbRedef}) we find that the value of the potential at the
minimum is:
\bea
\langle V_{\mc{O}(\vo^{-3})}\rangle&=&\frac{3}{4\langle s\rangle}\frac{W_0^2}{\tau_b^{9/2}}
\left\{-\frac{\left[\ln\left(\tau_b^{3/2}/W_0\right)\right]^{1/2}}{2 a^{3/2}} \right. \nn \\
&& \left. +\frac{8}{9\alpha}
\left(\frac{h b B}{W_0}\right)^2 \left(1- h b \beta\right)\, e^{-2 b \langle s\rangle}\tau_b^{3 (1- h b \beta)}
\left(1-\frac{3\alpha W_0}{2 h b B} \,\beta\,e^{b \langle s\rangle} \tau_b^{-\frac 32 (1- h b \beta)}\right)\right\}. \nn
\label{VO3minRedef}
\eea
In order to find a Minkowski vacuum we need therefore to perform the following tuning:
\be
\frac{\left[\ln\left(\tau_b^{3/2}/W_0\right)\right]^{1/2}}{2 a^{3/2}}=
\frac{8}{9\alpha}
\left(\frac{h b B}{W_0}\right)^2 \left(1-h b \beta\right)\, e^{-2 b \langle s\rangle}\tau_b^{3 (1-h b \beta)}
\left(1-\frac{3\alpha W_0}{2 h b B} \,\beta\,e^{b \langle s\rangle} \tau_b^{-\frac 32 (1- h b \beta)}\right)\,,
\label{tuningRedef}
\ee
which can be rewritten as:
\be
X^2- \alpha\beta X = C\,,
\label{quadratic}
\ee
where:
\be
X\equiv \frac 23 \frac{h b B}{W_0}\, e^{-b\langle s\rangle}\tau_b^{\frac 32 (1-h b \beta)}
\quad \text{and}\quad C\equiv\frac{\alpha}{4 (1- h b \beta)a^{3/2}}\left[\ln\left(\tau_b^{3/2}/W_0\right)\right]^{1/2}\,.
\ee
The solution to (\ref{quadratic}) is:
\be
X=\frac{\alpha\beta}{2}\left(1\pm \sqrt{1+\frac{4 C}{(\alpha\beta)^2}}\right)\,.
\label{solquad}
\ee
Given that, for $\vo\gg 1$ and $\beta < 1$ \footnote{In the case of a pure $SU(N)$ gauge theory $b_0=3 N$, and so
$\beta=N/(4\pi)$.}, the quantity $4 C/(\alpha\beta)^2$ scales as:
\be
\frac{4 C}{(\alpha\beta)^2}\sim  \frac{\mc{O}(1)}{(1- h b \beta)} \frac{\sqrt{\ln\vo}}{\beta^2}\gg 1\,,
\ee
it is crucial to consider separately the two cases $(1-h b \beta)>0$ and $(1-h b \beta)\leq 0$.
In the case $(1-h b \beta)\leq 0$ there is no real solution to (\ref{quadratic}),
preventing the possibility to obtain a dS vacuum.
However, given that $b=6\pi/b_0$, from the expression (\ref{beta}) for $\beta$,
we realise that $h b \beta=1/2$, and so the quantity $(1-h b \beta)=1/2$ is always positive.
Therefore the discriminant in (\ref{solquad}) is also always positive,
and so we correctly obtain two real solutions, one positive and one negative.
We shall discard the negative solution and keep the positive one which takes the form
(writing $\tau_b\simeq \vo^{2/3}$):
\be
B \simeq \underset{\mc{O}(1)}{\underbrace{\left(\frac{3\langle\tau_s\rangle^{1/4}}{4 h b}\sqrt{\frac{\alpha}{a}}\right)}}
\,\left(\frac{\langle\vo\rangle}{W_0}\right)^{\frac{b}{a} \left(\frac{2}{\zeta}\right)^{2/3}-(1-h b \beta)}
\,\frac{W_0^{h b \beta}}{\sqrt{1- h b \beta}}\,.
\ee
Notice that this solution, in the limit $\beta\to 0$, correctly
reduces to the one we found for the case with no field redefinition (see (\ref{sol})).
Hence the order of magnitude of $B$ needed to get a Minkowski vacuum depends again on
the Euler number and the choice of parameters $a$ and $b$.

\section{De Sitter LVS for non-perturbative effects only at the singularity}
\label{OnlyNPquiver}

Let us now consider what happens in the simplest set-up with no D7/E3 non-perturbative effects
but only with D3/E(-1) non-perturbative effects at the quiver locus.
This corresponds to the case analysed in section \ref{NPsing} without the $T_s$ superfield.

\subsection*{Case without field redefinition}

The scalar potential then takes the form
(neglecting the prefactor and writing $\langle s \rangle=g_s^{-1}$):
\be
V= \frac{2}{\alpha}(h b B)^2 \frac{e^{-2 b/g_s}}{\vo}+\frac 34 \frac{\zeta}{g_s^{3/2}} \frac{W_0^2}{\vo^3}\,.
\ee
If $\zeta\geq 0$ there is no extremum whereas is $\zeta<0$ the potential develops a maximum but not a minimum.
We need therefore to consider again string loop effects which could come either
from the visible sector cycle $\tau_{\rm vs}$ fixed by the interplay of D-terms and $g_s$ corrections:
\be
V_{\rm loop}^{({\rm vs})}\sim\frac{c_{\rm loop}^{({\rm vs})} W_0^2}{\vo^3 \sqrt{\langle\tau_{\rm vs}\rangle}}\,,
\label{loopvs}
\ee
or from the large cycle $\tau_b$ as in (\ref{Looplarge}).
In order to get a minimum, we need to consider both (\ref{loopvs})
and (\ref{Looplarge}) with $c_{\rm loop}^{({\rm b})}>0$. The potential becomes:
\be
V= \lambda_1 \frac{e^{-2 b/g_s}}{\vo}-\lambda_2 \frac{W_0^2}{\vo^3}
+\frac{c_{\rm loop}^{({\rm b})} W_0^2}{\vo^{10/3}}\,,
\ee
where:
\be
\lambda_1\equiv\frac{2}{\alpha}(h b B)^2>0\,,\qquad\lambda_2\equiv
\frac{c_{\rm loop}^{({\rm vs})}}{\sqrt{\langle\tau_{\rm vs}\rangle}}-\frac 34 \frac{\zeta}{g_s^{3/2}}>0 \,.
\ee
The equations $V'=0$, $V''>0$ and $\langle V\rangle =0$ admit a solution at:
\be
\langle\vo\rangle=\sqrt{\frac{\lambda_2}{7\lambda_1}}\,W_0\,e^{b/g_s}\quad\text{for}\quad
c_{\rm loop}^{({\rm b})}=\frac{6}{\lambda_1^{1/6}}\left(\frac{\lambda_2}{7}\right)^{7/6}\,W_0^{1/3}\,e^{b/(3 g_s)}\sim\mc{O}(\vo^{1/3})\,,
\label{tuningNogeom}
\ee
corresponding to a Minkowski minimum at exponentially large volume.
Notice that if $\zeta\leq 0$ then $\lambda_2$ is always positive whereas for $\zeta>0$ we need to tune
$c_{\rm loop}^{({\rm vs})}> g_s^{-3/2}\sqrt{\langle\tau_{\rm vs}\rangle}$
in order to get a positive $\lambda_2$. However this tuning would bring us to a
regime where we cannot trust the perturbative expansions since:
\be
\epsilon_{\rm loop}^{({\rm vs})}\sim\frac{c_{\rm loop}^{({\rm vs})}}{\tau_{\rm vs}}>\frac{g_s^{-3/2}}{\sqrt{\tau_{\rm vs}}}
\sim g_s^{-1}\gg 1\qquad\text{for}\qquad \tau_{\rm vs}\sim g_s^{-1}\sim\mc{O}(10)\,.
\ee
Hence we managed to obtain a dS LVS only for $\zeta\leq 0$ given that in this case no tuning is required on $c_{\rm loop}^{({\rm vs})}$
whereas the tuning on $c_{\rm loop}^{({\rm b})}$ does not ruin the perturbative expansion since from (\ref{tuningNogeom}) we realise that:
\be
\epsilon_{\rm loop}^{({\rm b})}\sim\frac{c_{\rm loop}^{({\rm b})}}{\vo^{2/3}}\sim \mc{O}(\vo^{-1/3})\ll 1
\qquad\text{for}\qquad \vo\gg 1\,.
\ee

\subsection*{Case with field redefinition}

The scalar potential then takes the form
(neglecting the prefactor and writing $\langle s \rangle=g_s^{-1}$ and $h b \beta= 1/2$):
\be
V= \frac{2}{\alpha}(h b B)^2 \frac{e^{-2 b/g_s}}{\vo^2}
   -3 B W_0 \frac{e^{-b/g_s}}{\vo^{5/2}}+\frac 34 \frac{\zeta}{g_s^{3/2}} \frac{W_0^2}{\vo^3} \,.
\ee
If $\zeta\leq 0$ there is just a maximum without any minimum. On the other hand, for positive $\zeta$
the equations $V'=0$, $V''>0$ and $\langle V\rangle =0$ have a solution at:
\be
\langle\vo\rangle=\left(\frac{3\alpha}{4 B (h b)^2}\right)^2 W_0^2\,e^{2b/g_s}\quad\text{for}\quad
g_s^{-3/2}=\frac{3 \alpha}{2 \zeta (h b)^2}\,,
\label{Tuning}
\ee
corresponding to a Minkowski minimum at exponentially large volume. However given that $\alpha$, $\zeta$, $h$
and $b$ are all $\mc{O}(1)$ numbers, (\ref{Tuning}) shows that it is hard to tune $g_s$ to get a dS vacuum
and still be in the perturbative regime where $g_s\lesssim 0.1$. Hence, in order to get a trustable vacuum,
we need again to consider the effect of string loop corrections.

In the case $\zeta=0$, we need to consider only (\ref{loopvs})
whereas we can neglect (\ref{Looplarge}) due to the volume suppression.
In fact, now the dS minimum would be located at:
\be
\langle\vo\rangle=\left(\frac{3\alpha}{4 B (h b)^2}\right)^2 W_0^2\,e^{2b/g_s}\quad\text{for}\quad
\frac{c_{\rm loop}^{({\rm vs})}}{\sqrt{\langle\tau_{\rm vs}\rangle}}= \frac{9 \alpha}{8 (h b)^2}\lesssim \mc{O}(1)\,,
\ee
in a region where we can still trust the perturbative expansion since:
\be
\epsilon_{\rm loop}^{({\rm vs})}\sim \frac{c_{\rm loop}^{({\rm vs})}}{\tau_{\rm vs}}\lesssim
g_s^{-1/2}\ll 1 \qquad\text{for}\qquad\tau_{\rm vs}\sim g_s^{-1}\sim\mc{O}(10)\,.
\ee
The situation for $\zeta\neq 0$ is more involved since, in order to get a minimum, we need also to
consider the loop correction (\ref{Looplarge}) coming from the large four-cycle.
We can set again $c_{\rm loop}^{({\rm vs})}= 0$ without loss of generality since the loop corrections
coming from the small cycle have the same volume scaling as the $\alpha'$ corrections and,
as we have seen before, they are always subdominant in the regime where we can trust the perturbative expansion.
The equations $V'=0$, $V''>0$ and $\langle V\rangle =0$ admit a solution only for $\zeta<0$ since
for positive $\zeta$ the loops can never modify the leading order dynamics due to their subleading volume scaling.
This dS LVS vacuum is located at:
\be
\langle\vo\rangle\simeq \mc{O}(1) \frac{W_0^2}{g_s^{3/4}}\,e^{2 b/g_s}\quad\text{for}\quad
c_{\rm loop}^{({\rm b})}\simeq \mc{O}(1) \frac{W_0^{2/3}}{g_s^{7/4}}\,e^{2 b/(3 g_s)}\sim\mc{O}(\vo^{1/3})\,, \nn
\ee
in a region where we can still trust the perturbative expansion. Notice that again, as in the case
without field redefinition, there is no minimum for $\zeta>0$.

\end{document}